\documentclass[10pt,preprint2]{aastex}

\def\insertfigure{\plotone}

\def\insertthreefigures#1#2#3{\plottwo{#1}{#2}\\\plotone{#3}}

\def\ion#1#2{\mbox{\rm #1\sc #2}}

\def\HII{{\ion{H}{ii}}}

\def\dim#1{\mbox{\,#1}}
\def\figdir{.}
\def\placefig#1{#1}

\begin{document}


\title{Secondary CMB Anisotropies from Cosmological Reionization}
\author{Nickolay Y.\ Gnedin}
\affil{CASA, University of Colorado, Boulder, CO 80309;
e-mail: \it gnedin@casa.colorado.edu}
\and
\author{Andrew H.\ Jaffe}
\affil{CfPA, University of California, Berkeley, CA 94720;
e-mail: \it jaffe@cfpa.berkeley.edu}

\begin{abstract}
  We use numerical simulation of cosmological reionization to calculate
  the secondary CMB anisotropies in a representative flat low density
  cosmological model. We show that the kinetic Sunyaev-Zel'dovich effect
  (scattering off of moving electrons in the ionized intergalactic
  medium) is dominated by the nonlinear hydrodynamic and gravitational
  evolution of the density and velocity fields, rather than the detailed
  distribution of the ionization fraction (``patchy reionization'') on
  all angular scales.  Combining our results with the recent calculation
  of secondary CMB anisotropies by Springel et al., we are able to
  accurately predict the power spectrum of the kinetic SZ effect on
  almost all angular scales.
\end{abstract}


\section{Introduction}

The reionization of the intergalactic medium (hereafter, IGM) stands out
as one of the most important physical processes that have taken place in
the early universe. Recent observational progress measuring the
angular spectrum of anisotropies in the Cosmic Microwave Background radiation
(hereafter CMB) on sub-degree angular scales (de Bernardis et al.\ 2000;
Hanany et al.\ 2000), and in obtaining the absorption spectra of the
most distant quasars (Stern et al.\ 2000; Zheng et al.\ 2000; Fan et
al.\ 2000a) allows us to limit the redshift of reionization to somewhere
between 6 and 30 (Griffiths, Barbosa, \& Liddle 1998; Bond \& Jaffe
1999; Tegmark \& Zaldarriaga 2000). Combined with recent theoretical and
numerical breakthroughs in our understanding how reionization proceeds
in the inhomogeneous universe (Giroux \& Shapiro 1996; Haiman \& Loeb
1997; Gnedin \& Ostriker 1997; Madau, Haardt, \& Rees 1999;
Miralda-Escude\'e, Haehnelt, \& Rees 1999; Ciardi et al.\ 1999; Valageas
\& Silk 1999; Chiu \& Ostriker 2000; Gnedin 2000), these data finally
place a study of reionization on much more solid footing.

Among the many effects of reionization, the secondary CMB anisotropies
have been the subject of active study for a long time, beginning with
the pioneering works of Ostriker \& Vishniac (1986) and Vishniac (1987)
and up to the most recent careful investigations (Hu \& White 1996;
Aghanim et al. 1996; Knox, Scoccimarro, \& Dodelson 1998; Gruzinov \& Hu
1998; Jaffe \& Kamionkowski 1998; Peebles \& Juszkiewicz 1998; Haiman \&
Knox 1999; Hu 2000; Bruscoli et al.\ 2000; da Silva et al.\ 2000;
Refregier et al.\ 2000; Seljak, Burwell, \& Pen 2000; Springel, White,
\& Hernquist 2000).  However, most of the recent studies have not
benefited from the recent progress on reionization, and were based on
over-simplified ad hoc models.  In addition, many of these studies only
considered the formerly favored ``Standard'' Cold Dark Matter (CDM)
model. However, there appears finally to be convergence toward a low
density CDM model with the cosmological constant (the so-called
CDM+$\Lambda$ model; c.f.\ White, Scott, \& Pierpaoli 2000; Tegmark \&
Zaldarriaga 2000; Bridle et al.\ 2000; Hu et al.\ 2000; Jaffe et al.\ 
2000), and thus it would make sense to calculate the secondary
anisotropies for this currently most favored model.

In this paper we reconsider the effect of reionization on the CMB, using
the latest numerical simulations of reionization (Gnedin 2000). By
combining our numerical results with analytical calculations and with
recent numerical calculation by Springel et al.\ (2000) on larger
angular scales, we are able to extend the range of angular scales over
which predictions are reliable to over more than six orders of
magnitude. We also take special care to make sure that our results
are not contaminated by numerical artifacts, the dominant of which is
the periodicity of the simulation boundary conditions.

\section{Theory}

\subsection{Secondary CMB Anisotropies}


Thomson scattering of CMB photons by ionized gas is usually
called the Sunyaev-Zel'dovich (SZ) effect, which is
often divided into ``thermal'' and ``kinetic'' SZ effects.
The ``thermal'' SZ effect is the upscattering of
CMB photons after compton scattering by hot gas (most pronounced in
clusters of galaxies).
The ``kinetic'' SZ effect is the temperature fluctuation induced by bulk
motions of the gas. Both effects play a role in generating secondary
CMB anisotropies during and after reionization.

The fractional
temperature perturbation in the direction ${\hat\theta}$ induced by bulk
motions (``kinetic effect'') is
\begin{equation}
  {\Delta T \over T}({\vec\theta}) =
  -\int_0^{\eta_0} \, n_e \sigma_T e^{-\tau}
  [{\hat\theta} \cdot {\vec v}(w{\hat\theta};w)] a(w)\; dw,
  \label{eq:startingpoint}
\end{equation}
where $n_e$ is the electron density along the line of sight, ${\vec
  v}({\vec w};w)$ is the bulk peculiar velocity at position ${\vec w}$ at a
conformal time $\eta_0-w$, $\sigma_T$ is the Thomson cross section, 
$\tau$ is the optical depth from us to $w$, and the subscript 0 denotes a 
quantity at the current moment.
The factor of $a(w)$ arises
because the physical time is $dt = a(w) dw$.  Note that ${\hat\theta}$
represents a three-dimensional unit vector along the line of sight,
whereas ${\vec\theta}$ will refer to a dimensionless two-dimensional
vector in the plane perpendicular to the line of sight---i.e., for
directions near ${\hat z}$, ${\hat\theta} = (\theta_1, \theta_2,
\sqrt{1-\theta_1^2-\theta_2^2}) \simeq (\theta_1, \theta_2, 1)$, whereas
${\vec\theta}=(\theta_1,\theta_2,0)$.  

It is convenient to rewrite equation (\ref{eq:startingpoint}) in the
following dimensionless form:
\begin{equation}
  {\Delta T \over T}({\vec\theta}) = -\tau_0
  \int_0^{\eta_0} \chi e^{-\tau}
  ({\hat\theta} \cdot {\vec v\over c}) {dw\over a^2 \eta_0},
\end{equation}
where $\tau_0 \equiv c\eta_0\sigma_T (\bar{n}_{\rm H,0}+2\bar{n}_{\rm He,0})$
and
$$
        \chi \equiv {n_e\over \bar{n}_{\rm H}+2\bar{n}_{\rm He}}
        = x_e (1+\delta).
$$
Here $x_e$ is the ionization fraction, $\delta$ is cosmic overdensity, and
$\bar{n}_{\rm H}$ and $\bar{n}_{\rm He}$ are the mean hydrogen and helium 
number densities.

Analogously, the ``thermal'' SZ effect is described by the following
integral (for observations in the Rayleigh-Jeans regime):
\begin{equation}
  {\Delta T \over T}({\vec\theta}) = 
  -2\tau_0\int_0^{\eta_0} \chi e^{-\tau}
  {k_B(T_g-T_0)\over m_ec^2} {dw\over a^2\eta_0},
  \label{thermalsz}
\end{equation}
where $T_g$ and $T_0$ are the gas and the CMB temperatures 
respectively.

For a particular realization, we can simply calculate integral 
(\ref{eq:startingpoint}) or (\ref{thermalsz}) along each
line of sight and make a map of temperature anisotropy. In this paper we
use numerical simulation of cosmological reionization (described in detail
in the following section) to directly calculate the anisotropy map
on the patch on the sky subtended by the computational box. 

Often however it is just 
the two-point correlation function of the anisotropies, 
$C(\theta)$, or its spherical-harmonic transform,
the power spectrum, $C_\ell$, which are of interest. In this case
we can directly calculate the correlation function from equation
(\ref{eq:startingpoint}),
\begin{equation}
        C(\theta) \equiv \langle {\Delta T \over T}({\vec\theta}_1)
        {\Delta T \over T}({\vec\theta}_2) 
        \rangle_{\cos\theta={\vec\theta}_1\cdot{\vec\theta}_2},
\end{equation}  
which can be reduced to the following form:
$$
        C(\theta) = \tau_0^2 
        \int_0^{\eta_0} {dw_1\over a_1^2\eta_0} e^{-\tau_1}
        \int_0^{\eta_0} {dw_2\over a_2^2\eta_0} e^{-\tau_2}
        \phantom{AAAA}
$$
\begin{equation}
        \phantom{AAAA}
        \times C_{ee}(\eta_1,\eta_2,\theta) C_{vv}(\eta_1,\eta_2,\theta),
        \label{corfuns}
\end{equation}
where
$C_{ee}$ and $C_{vv}$ are the electron density and velocity
correlation functions,
\begin{eqnarray}
  \label{eq:corrfunctions}
  C_{ee}(\eta_1, \eta_2, \theta) & \equiv &
  \langle \chi({\vec r}_1, \eta_1) \chi({\vec r}_2, \eta_2)\rangle,
  \nonumber\\
  C_{vv}(\eta_1, \eta_2, \theta) & \equiv &
  \langle {\hat\theta}_1{\vec v\over c}({\vec r}_1, \eta_1) \cdot
        {\hat\theta}_2{\vec v\over c}({\vec r}_2, \eta_2)\rangle,
\end{eqnarray}
with $\cos\theta=({\vec r}_1\cdot{\vec r}_2)/(r_1r_2)$,
provided we make a usual assumption (c.f.\ Bruscoli et al.\ 1999) 
that the electron density and velocity are uncorrelated, i.e.,
\begin{equation}
 \label{eq:cev}
  C_{ev}(\eta_1, \eta_2, \theta) \equiv
  \langle \chi({\vec r}_1, \eta_1) {\hat\theta}_2\cdot{\vec v\over c}
        ({\vec r}_2, \eta_2)\rangle = 0.
\end{equation}
We will return to the accuracy of this assumption below.

\subsection{Terminology for Secondary Anisotropies}

In reality we have to deal with only one universe, and thus we can only
observe the total CMB anisotropies given by equation
(\ref{eq:startingpoint}), generically labeled as the kinetic
Sunyaev-Zel'dovich effect.  Historically, this was first investigated by
restricting to a much simpler, unphysical situation: homogeneous
reionization and the linear evolution of velocity and density fields.
This is the ``Ostriker-Vishniac effect'' (OV; Ostriker \& Vishniac 1986;
Jaffe \& Kamionkowski 1998), also sometimes called the ``Vishniac
effect.'' In this case, we observe the autocorrelation of the total gas
density and velocity fields, related to the power spectrum of density
fluctuations by linear theory.

In this work, we will further distinguish between the ``Linear
Ostriker-Vishniac Effect'' (LOV) and the ``Nonlinear Ostriker-Vishniac
effect'' (NLOV). The former is the usual calculation of OV, using the
linear relationship between velocity and density. This comes into the
needed calculation of the four-point function $\langle v\delta v\delta
\rangle$, in which it is assumed that (1) the linear relationship
between $v$ and $\delta$ holds; and (2) that the four-point function is
appropriate for an underlying Gaussian density field, which only holds
in the fully linear regime (and only with Gaussian initial conditions).
The NLOV assumes a homogeneous ionization fraction, but the full
nonlinear gas density and velocity fields.

Recently, the LOV calculations have been complimented by consideration of
the distribution of the ionized hydrogen in the universe as it undergoes
reionization, so-called ``patchy'' or ``inhomogeneous'' reionization.
Workers in the field have considered models for the distribution of the
ionization fraction in the linear regime (e.g., Knox, Scoccimarro, \&
Dodelson 1998; Gruzinov \& Hu 1998), as well as full simulations or
semianalytic models of the ionized gas, as in this work (see also
Bruscoli et al.\ (2000), Benson et al.  (2000), and Springel et al
(2000)).

This separation is obviously somewhat artificial.  Nevertheless, for academic purposes, we can denote those two effects in the
following fashion:
\begin{equation}
  \left.\Delta T \over T\right|_{OV} =
  -\tau_0\int_0^{\eta_0} \bar{x}_e (1+\delta) e^{-\tau}
  ({\hat\theta} \cdot {\vec v\over c}) {dw\over a^2\eta_0}
  \label{oveff}
\end{equation}
for the NL Ostriker-Vishniac effect, and
\begin{equation}
  \left.\Delta T \over T\right|_{PR} =
  -\tau_0\int_0^{\eta_0} \, (x_e-\bar{x}_e) (1+\delta) e^{-\tau}
  ({\hat\theta} \cdot {\vec v\over c}) {dw\over a^2\eta_0}
  \label{preff}
\end{equation}
for patchy reionization.

Thus, the NL Ostriker-Vishniac effect describes anisotropies generated
in the universe in which the ionization fraction is homogeneous in space
(``homogeneous reionization''), and patchy reionization includes the
rest of the anisotropies. We again emphasize that this separation is
artificial and unphysical, and we adopt it purely for historical
reasons, as the two effects were often considered to be ``distinct''
effects.

Using the above definition, we can also define the electron density
correlation functions due to NL Ostriker-Vishniac and patchy reionization
effects,
\begin{eqnarray}
  \label{eq:ovprcf}
  C_{ee}^{OV}(\eta_1, \eta_2, \theta) & \equiv &
  \bar{x}_e^2 \langle \delta({\vec r}_1, \eta_1) \delta({\vec r}_2, \eta_2)
  \rangle,
  \nonumber\\
  C_{ee}^{PR}(\eta_1, \eta_2, \theta) & \equiv &
  \langle \Delta\chi({\vec r}_1, \eta_1) \Delta\chi({\vec r}_2, \eta_2)\rangle,
\end{eqnarray}
where $\Delta\chi\equiv (x_e-\bar{x}_e)(1+\delta)$. The total correlation 
function $C_{ee}$ is then a sum of these two functions plus the 
cross-correlation,
$$
        C_{ee} = C_{ee}^{OV} + C_{ee}^{PR} + 2C_{ee}^{OV-PR},
$$
where
\begin{equation}
  C_{ee}^{OV-PR}(\eta_1, \eta_2, \theta) \equiv
  \bar{x}_e \langle \delta({\vec r}_1, \eta_1) \Delta\chi({\vec r}_2, \eta_2)
  \rangle.
  \label{ovprcc}
\end{equation}

\section{Method}

\subsection{Simulation}

We use a cosmological simulation of reionization
reported in Gnedin (2000). The simulation includes 3D radiative
transfer (in an approximate implementation) and other physical ingredients
required for modeling the process of cosmological reionization. 

The simulation of a representative CDM+$\Lambda$ cosmological 
model\footnote{With the following cosmological parameters: $\Omega_0=0.3$,
$\Omega_\Lambda=0.7$, 
$h=0.7$, $\Omega_b=0.04$, $n=1$, $\sigma_8=0.91$, where the amplitude and
the slope of the primordial spectrum are fixed by the COBE and cluster-scale
normalizations.}
was performed in a comoving box with the size of $4h^{-1}\dim{Mpc}$ with the
mass resolution of $5\times10^5\dim{M}_\odot$ in baryons and the comoving
spatial resolution of $1h^{-1}\dim{kpc}$.

The simulation was stopped at $z=4$ because at this time the rms density 
fluctuation in the computational box is about 0.25, and at later times the box
ceases to be a representative region of the universe.

As was noted in Gnedin (2000), this simulation is still insufficient to give 
the fully converged numerically result. As we show below, the computational 
box size of this simulation is not large enough to allow for accurate
computation of the CMB anisotropies, as perturbations at $z<4$ contribute
about 20\% of the signal. We therefore must emphasize here that our 
calculations are valid only on a semi-qualitative basis, within a factor of
1.5-2, and a still larger simulation is required to accurately model all the
relevant scales present in the problem. But since the observations of the 
secondary anisotropies are some years away, theorists have time to improve
upon their models and perform larger, highly accurate simulations.

Our simulation assumes that all cosmological reionization occurs via
radiation from {\em stellar} sources, with star formation parameterized
by the phenomenological Schmidt law as discussed in Gnedin (2000). We
ignore alternative and complimentary scenarios in which the bulk of
reionization is produced by Active Galactic Nuclei. In light of recent
high-redshift quasar counts (e.g., Fan et al.\ 2000b), a scenario in which
the universe is reionized by bright optically selected QSOs seems
unlikely in any event. There remains however the possibility that 
low brightness AGNs contributed significantly (if not dominantly) to the
reionization of the universe. In the latter case however they will be clustered
on scales similar to the stellar sources, and thus will not result in a
qualitatively different reionization scenario, although the epoch of
reionization in this case is less well-determined.

\subsection{Numerical Issues}

Given a cosmological simulation, we are now faced with the task of
calculating the secondary CMB anisotropies accurately from the
simulation. Since perturbations are generated over a considerable
redshift range, ideally we would like to have a simulation with the box
size that extends from $z=0$ until $z\sim30$ in one direction. However,
with currently feasible simulations this is impossible, and we utilize
the cubic box with the $4h^{-1}\dim{Mpc}$ side used in Gnedin (2000).

\def\capPB{
Cartoon showing the effect of the periodic box on the calculated CMB 
anisotropies. Right panels depict a piece of the universe along the
line of sight
 with \HII\ regions
in it, and left panels show the correlation function of any physical
quantity as a function of the distance $w$ along the line of sight.
The upper row depicts the actual Universe, with random positions
for \HII\ regions and the correlation function extending to large 
scales. The middle row shows the periodic universe as given in a cosmological
simulation. Finally, the lower row depicts our method of calculating the
anisotropies, with the periodic box randomly shifted and flipped in the 
subsequent images. Arrows in the last two rows show the orientation of the
computational box. For simplicity, we do not show random shift of the box
in the last row which is actually performed in our calculation in addition
to random flipping and transposing.
}
\placefig{
\begin{figure}
\insertfigure{\figdir/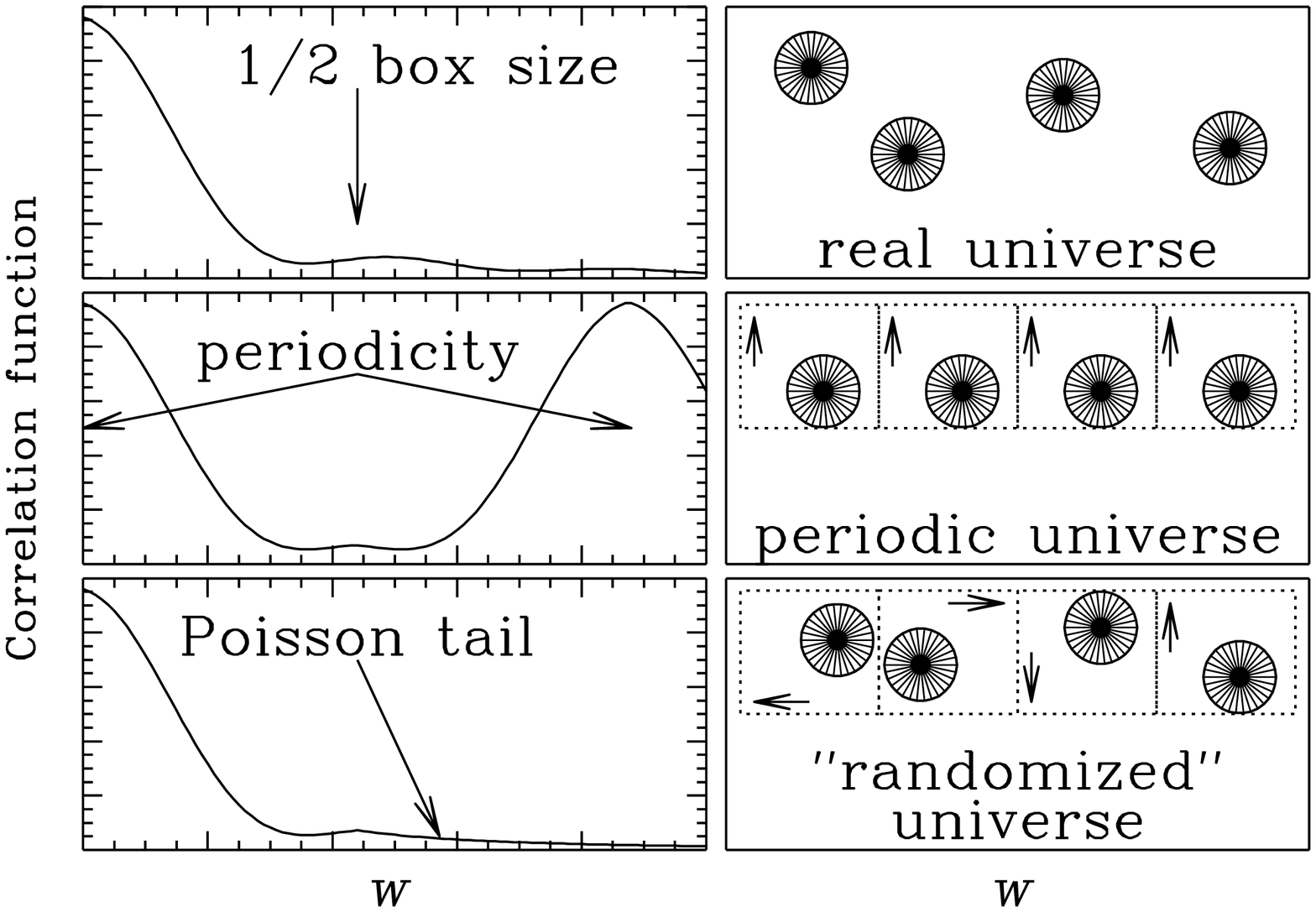}
\caption{\label{figPB}\capPB}
\end{figure}
} 
If we take the output of a simulation as is, the artificial
periodicity of the universe will amplify the anisotropies by a large
factor, simply because in real universe the signal will be averaged over
many randomly placed \HII\ regions, while in a simulated periodic
universe a photon, on its way along the line of sight,
will encounter the same structure over and over again.
This case is depicted in the first two rows of Figure \ref{figPB}. In
order to avoid this artificial amplification, we randomly flip and
transpose the computational box around any of its six edges, and in
addition shift it by a random distance in a random direction.  This
``randomization'' procedure ensures that the next periodic image of the
computational box is not correlated with the given image, and thus we
lose the correlations over the scales larger than half the size of the
computational box along the line-of-sight direction.
(In addition to the flipping, transposition and
shifting of the simulation box, we have also rotated it by a random
amount, but this has negligible effect).  This is sketched in the third
row of Fig.\ \ref{figPB}.  Thus, our procedure actually {\it
  underestimates\/} true anisotropies by ignoring large-scale signal.
However, we can add the missing large-scale signal using the linear
theory calculation, as we discuss below.

In addition, since we cannot unambiguously eliminate periodicity in the plane
of the sky, we restrict our image to the image of one periodic simulation
box, which corresponds to the angular size of 
$$
        \theta_{\rm max} = {L\over R(t_{\rm im})}
$$
where $L=4h^{-1}\dim{Mpc}$ is the comoving size of our computational box,
$R(t)$ is the comoving angular-diameter distance from time $t$ to the
current moment,
$$
        R = c\int_{t}^{t_0} {dt\over a(t)},
$$
and we choose to scale all the computational boxes along the line-of-sight
to the box taken at $a_{\rm im}\equiv a(t_{\rm im})=1/(1+z_{\rm im})=0.115$, 
the epoch just
before reionization which contributes most to the CMB anisotropies.
For our calculations, $\theta_{\rm max}\simeq60$ arcsec corresponding to a
spherical harmonic multipole of $\ell\simeq 10^4$.

Because in the simulation the bulk velocity of the computational box is
set to zero, whereas a volume with the size of our box in real universe
will have some non-vanishing bulk velocity, we add to each box a random
bulk velocity drawn from the Gaussian distribution with the dispersion
equal to the linear rms velocity on the scale of the computational box
at each redshift. Because this effect depends on the correlation of
density with velocity, our results on all angular scales depend on such
an appropriate realization of the velocities.

In summary, 
our procedure to calculate the CMB anisotropies from the cosmological
simulation with a periodic computational box can be described as follows:
\begin{enumerate}
\item We select the outputs from the simulation spaced with $\Delta a$ in
scale factor.
\item For each pair of outputs at $a_i$ and $a_{i+1}$, where index $i$ 
numbers the outputs, we calculate the number of computational boxes along
the line of sight between $a_i$ and $a_{i+1}$. 
\item Then for all boxes between
$a_i$ and $(a_i+a_{i+1})/2$ we fill them in with the simulation output
data at $a_i$, and the boxes between $(a_i+a_{i+1})/2$ and $a_{i+1}$ we fill
in with the data at $a_{i+1}$. 
\item Then each of thus-filled-in boxes (including
the original simulation boxes at $a_i$ and $a_{i+1}$) is randomized by a
random shift in the plane of the sky and by a random flip and transposition,
removing correlations on scales larger than half the box size. In addition,
each box is assigned a bulk velocity along the line-of-sight according to the
linear theory.
\item We repeat steps 2-4 for all pairs of outputs in the descending order
from the last output
at $z=4$ ($a=0.2$) until the start of the simulation at $\sim100$.
\item After arranging all the boxes, we calculate the CMB map on a
  square patch on the sky (corresponding to the image of our cubic
  computational box at $z=7.7$, the epoch when about half the box is
  ionized) with a given resolution (we achieve the highest resolution of
  $512^2$ pixels).
\end{enumerate}

This procedure allows us to construct an image on the sky. However, it 
is customary to present the CMB anisotropies on the sky in terms of the
power spectrum $C_\ell$, which is the spherical-harmonic transform of the
correlation function $C(\theta)$. In the small-angle limit, the 
spherical-harmonic transform reduces to the Hankel transform of zeroth order,
\begin{equation}
        C_\ell=2\pi\int_0^\infty d\theta\,\theta\,J_0(\theta\ell)C(\theta),
        \label{cellcf}
\end{equation}
where $J_0$ is the zero order Bessel function. However, because of the 
numerical errors and artificial ringing due to a finite box size, the power 
spectrum calculated from equation (\ref{cellcf}) is noisy. An alternative
method of calculating the power spectrum is a direct Fourier transform of the
CMB anisotropies on the sky,
\begin{equation}
        C_\ell=\theta_{\rm max}^2 \langle | \Delta_{\vec \ell} |^2 \rangle,
        \label{cellft}
\end{equation}
where $\theta_{\rm max}$ is the angular size of a square patch on the sky
(obviously, $\theta_{\rm max}\ll1$), and
$$
        \Delta_{\vec \ell} \equiv {1\over\theta_{\rm max}^2} 
	\int {\Delta T\over T}
        (\vec\theta) e^{\displaystyle i\vec\theta\cdot\vec\ell} d^2\theta
$$
is the Fourier transform of the CMB anisotropies.

\def\capPS{
Comparison of the power spectrum of the CMB anisotropies computed by equation
(\protect{\ref{cellcf}}) ({\it dotted line\/}), and by equation
(\protect{\ref{cellft}}) ({\it solid line\/}). In the latter case also shown
are the statistical error-bars for the computed power spectrum.
Two arrows show the
angular size of the computational box and the angular resolution of our
calculation respectively (for the $256^2$ pixelization).
}
\placefig{
\begin{figure}
\insertfigure{\figdir/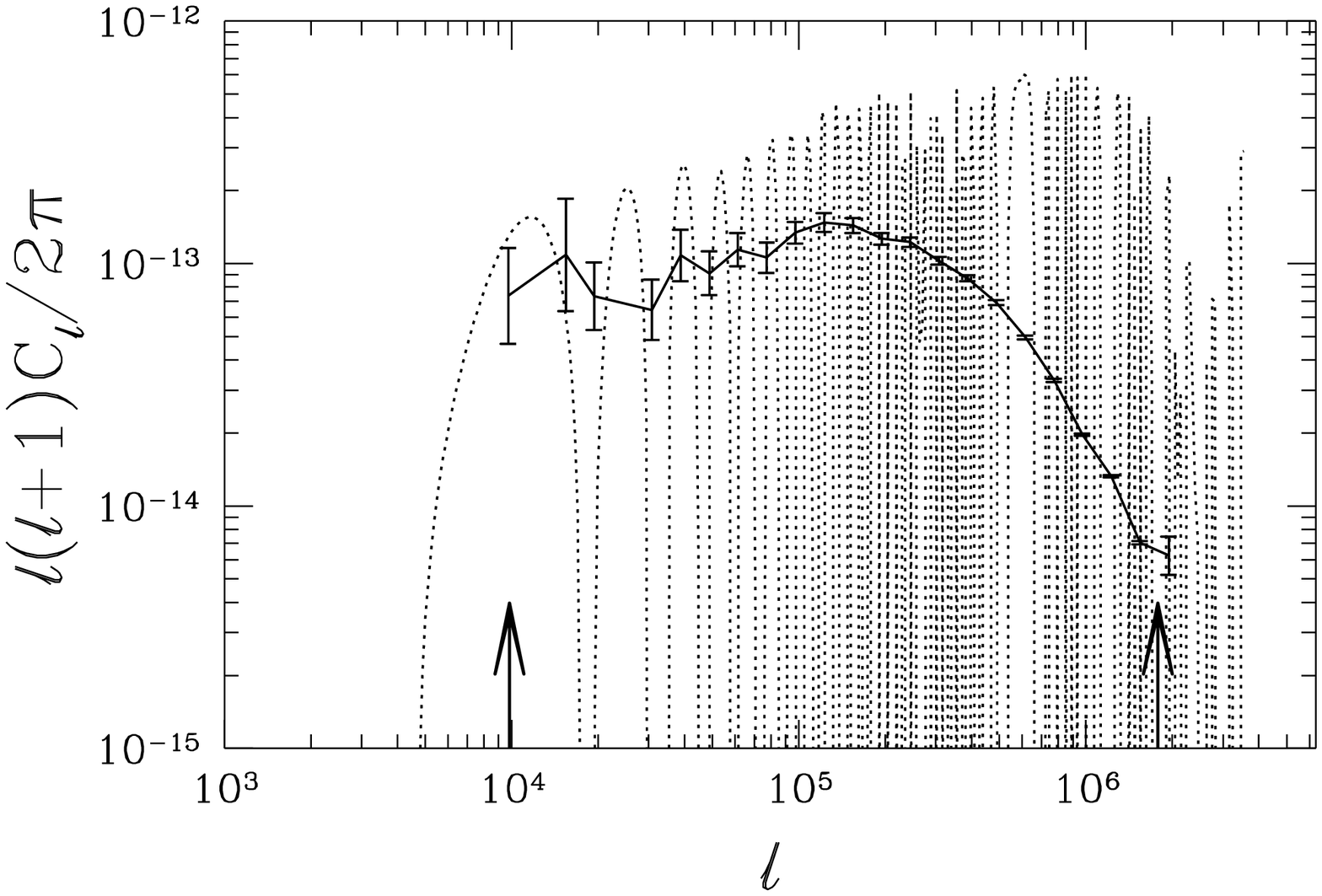}
\caption{\label{figPS}\capPS}
\end{figure}
}
Figure \ref{figPS} shows a comparison between the two methods for computing
the power spectrum $C_\ell$. One can see that the correlation function method
produces an extremely noisy estimate of the power spectrum, whereas the one
computed directly is well defined up to the maximal possible multipole
$$
        \ell_{\rm max}=\sqrt{2} \ell_{\rm Ny} = 
        \sqrt{2}N{\pi\over\theta_{\rm max}},
$$
where $\ell_{\rm Ny}$ is the Nyquist frequency of a square image on the sky
and $N$ is the number of pixels along one dimension in the image.

\def\capPC{
Corrected ({\it bold lines\/}) and uncorrected ({\it thin lines\/})
power spectra at four different angular resolutions: 
$64^2$ pixelization ({\it dotted lines\/}),
$128^2$ pixelization ({\it short-dashed lines\/}), 
$256^2$ pixelization ({\it long-dashed lines\/}), and
$512^2$ pixelization ({\it solid lines\/}).
Notice that our correction fully removes the effect of pixelization on small
scales.
}
\placefig{
\begin{figure}
\insertfigure{\figdir/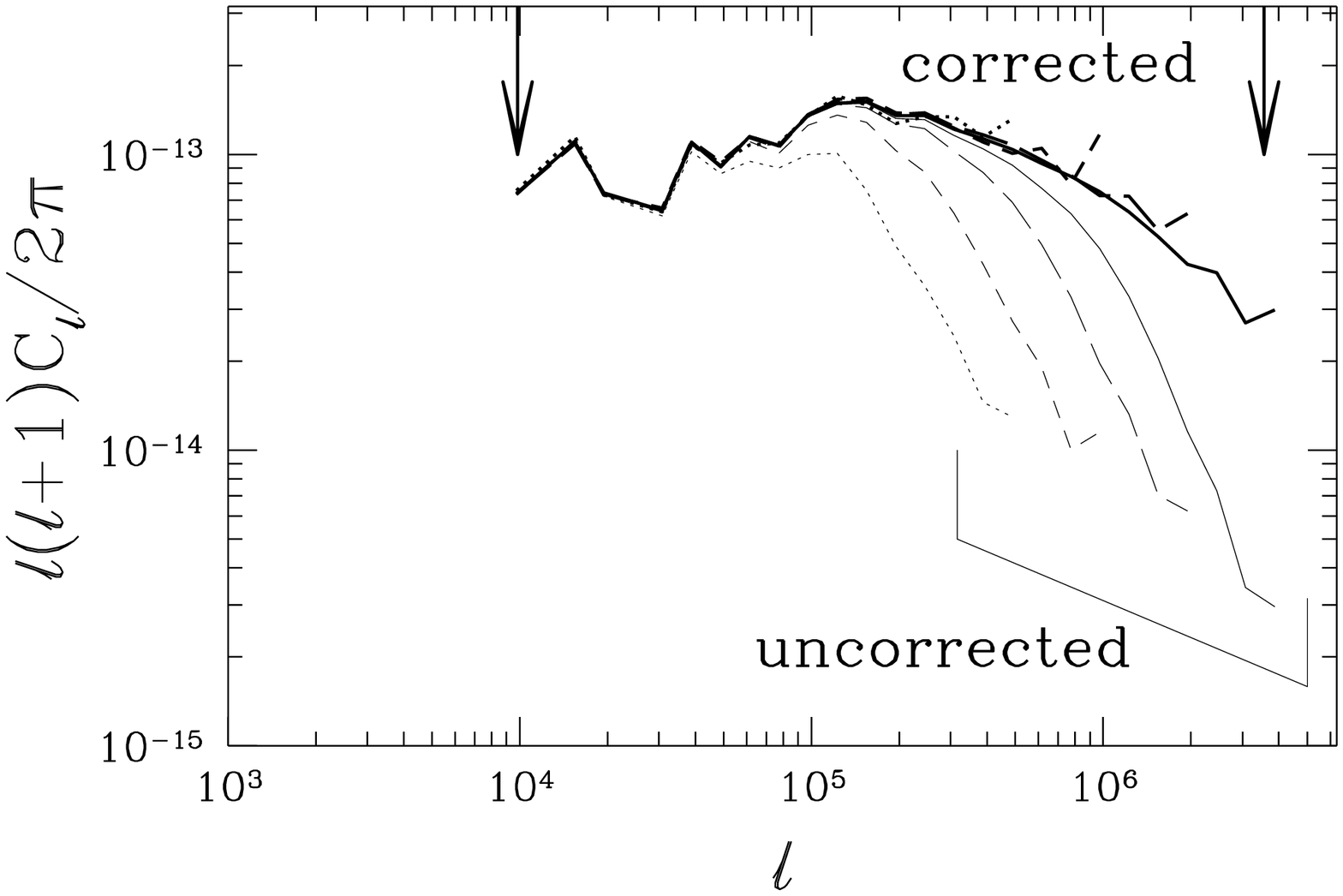}
\caption{\label{figPC}\capPC}
\end{figure}
} The process of pixelization - binning of the computed image on the
square patch on the sky of a given size - produces small-scale smearing
similar to the one produced by assigning density on a regular mesh from
a particle distribution in cosmological $N$-body simulations. However,
in our case the precise form of this smoothing cannot be computed from
the first principles, because the computational box subtends somewhat
different angular size at different redshifts. Instead, we correct for
the small-scale smearing empirically. Let $C_\ell$ be the true power
spectrum.  Since the smearing is caused by pixelization, the computed
power spectrum $\tilde C_\ell$ will be a product of the true power
spectrum and the window function $W$ which depends only on the ratio of
the multipole $\ell$ and the maximal possible multipole $\ell_{\rm max}$
for a given pixel size, or, in other words, on the product of $\ell$ and
the number of pixels along one dimension $N$. Thus,
$$
        \tilde C_\ell = W(\ell/\ell_{\rm max}) C_\ell,
$$
and if we take the ratio of the power spectra from two different 
pixelizations, this ratio will depend only on the window function,
\begin{equation}
        {\tilde C_{1, \ell}\over \tilde C_{2, \ell}} =
        {W(\ell/\ell_{1, \rm max})\over W(\ell/\ell_{2, \rm max})}.
        \label{winfunrat}
\end{equation}
Using equation (\ref{winfunrat}), we find the following approximate form
for the window function
\begin{equation}
        W(\ell/\ell_{\rm max}) = {\left[1+10/3(\ell/\ell_{\rm max})^2\right]^2
        \over 1+(\ell/\ell_{\rm max})^4}.
        \label{winfun}
\end{equation}
Figure \ref{figPC} shows the uncorrected and corrected power spectra for
four resolutions: $N=64$, $128$, $256$, and $512$. 
One can see that the corrected power
spectra agree with each other up to their respective $\ell_{\rm max}$. We also notice
that the power spectrum seems to continue as a power law with index -3 to the
smallest scales. Thus, we are unable to achieve convergence on small
angular scales, which implies that even with resolution of $512^2$ pixels
per 2.2 arcmin patch on the sky, there exists structure on the unresolved
scales. As we mentioned above, this structure is due to small-scale
density inhomogeneity in the ionized high density regions. Our simulation
has a dynamical range of 4000, and thus in principle
we can measure the power spectrum of the secondary anisotropies from our
simulation up to 
$l\sim 3\times10^7$ (compared to $l=3.5\times10^6$ for our highest $512^2$
resolution). However, making CMB maps with resolution higher than
$512^2$ is beyond the limit of computational resources available to us
(doing a $4000^2$ image would take more computer time than the
simulation itself consumed). We thus conclude that the cut-off in the
power spectrum at $l\sim3\times10^5$ reported in Bruscoli et al.\ (2000)
is not real and due to the finite angular resolution ($256^2$) of their
final CMB map.

\def\capAC{
Power spectra of the CMB anisotropies pixelized on the $256^2$ image computed
with two different values of the sampling 
frequency of numerical integration: 
$\Delta a=0.01$ ({\it dotted line\/}),
$\Delta a=0.005$ ({\it solid line\/}).
The dashed line shows a
calculation with no randomization, i.e., with the periodic universe,
which gives about a factor of 3-10 larger signal. Two arrows give the
angular size of the computational box and the angular resolution of our
calculation respectively (for the $256^2$ pixelization).
}
\placefig{
\begin{figure}
\insertfigure{\figdir/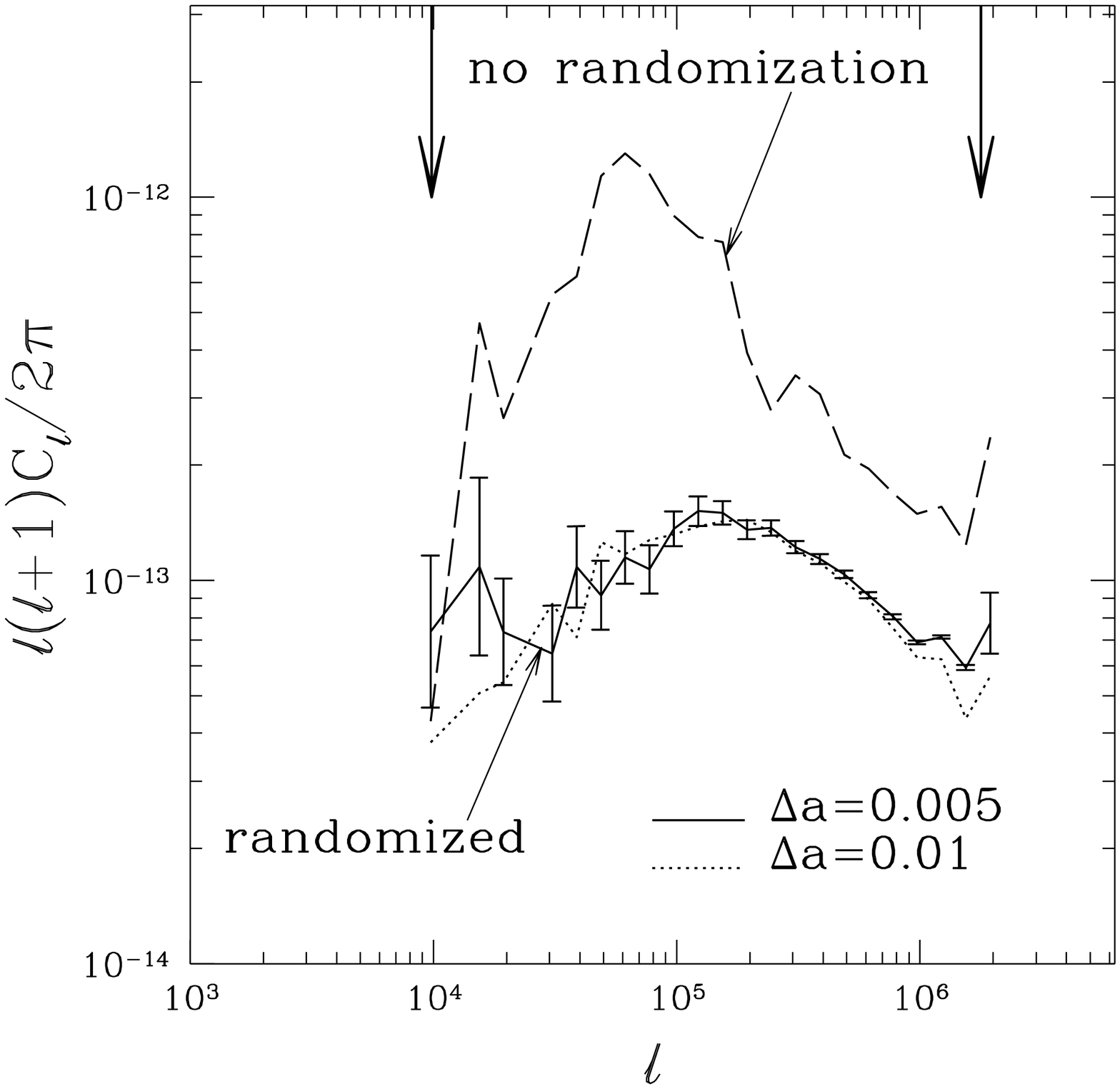}
\caption{\label{figAC}\capAC}
\vspace*{2in}
\end{figure}
}
The outputs from our simulation are only stored at discreet moments in time
(which we choose to parametrize with the cosmological scale factor
$a\equiv1/[1+z]$). Integration over the path of a photon is done
numerically using the outputs from the simulation, and thus the numerical
value of the integral depends on the interval $\Delta a$ between the
consequent outputs, converging to the exact value in the limit
$\Delta a\rightarrow0$. Of course, we cannot reach this limit as it would
require an infinite number of outputs, and thus we must ensure that our
choice of $\Delta a$ is sufficiently small to give an accurate value for the
integral over the photon path. Figure \ref{figAC} shows the power
spectra of the CMB anisotropies 
for two choices of $\Delta a$. Both values
give convergent
results for the CMB anisotropies, and
so we adopt $\Delta a=0.005$ as our choice for this parameter
(which is equivalent to having about 40 outputs until $z=4$). It also
becomes prohibitively expensive to calculate the integral with a smaller
$\Delta a$. We also show in Fig.\ \ref{figAC} with the long-dashed line
the CMB anisotropies calculated without randomization (i.e., in an
unrealistic periodic universe), which are off by up to a factor of 10.

We would like to note here that a recent similar analysis by Bruscoli
et al.\ (2000) 
did not use randomization in calculating
the CMB anisotropies, and by adopting a periodic universe, overestimated
the correct result by a factor of 3-10.

\def\capNC{
({\it a\/}). The rms CMB temperature anisotropy as a function of the 
final redshift of
integration. The solid line shows our simulation, which is stopped at
$z=4$, and the dotted line shows an extrapolation to $z=0$ assuming that
perturbations do not change in comoving coordinates. ({\it b\/}). 
A comparison
of the perturbation power spectrum as extrapolated to $z=0$ ({\it solid 
line\/}) and at $z=4$ ({\it dashed line\/}). The difference between the two 
curves is representative of the uncertainty of our calculation. 
}
\placefig{
\begin{figure}
\insertfigure{\figdir/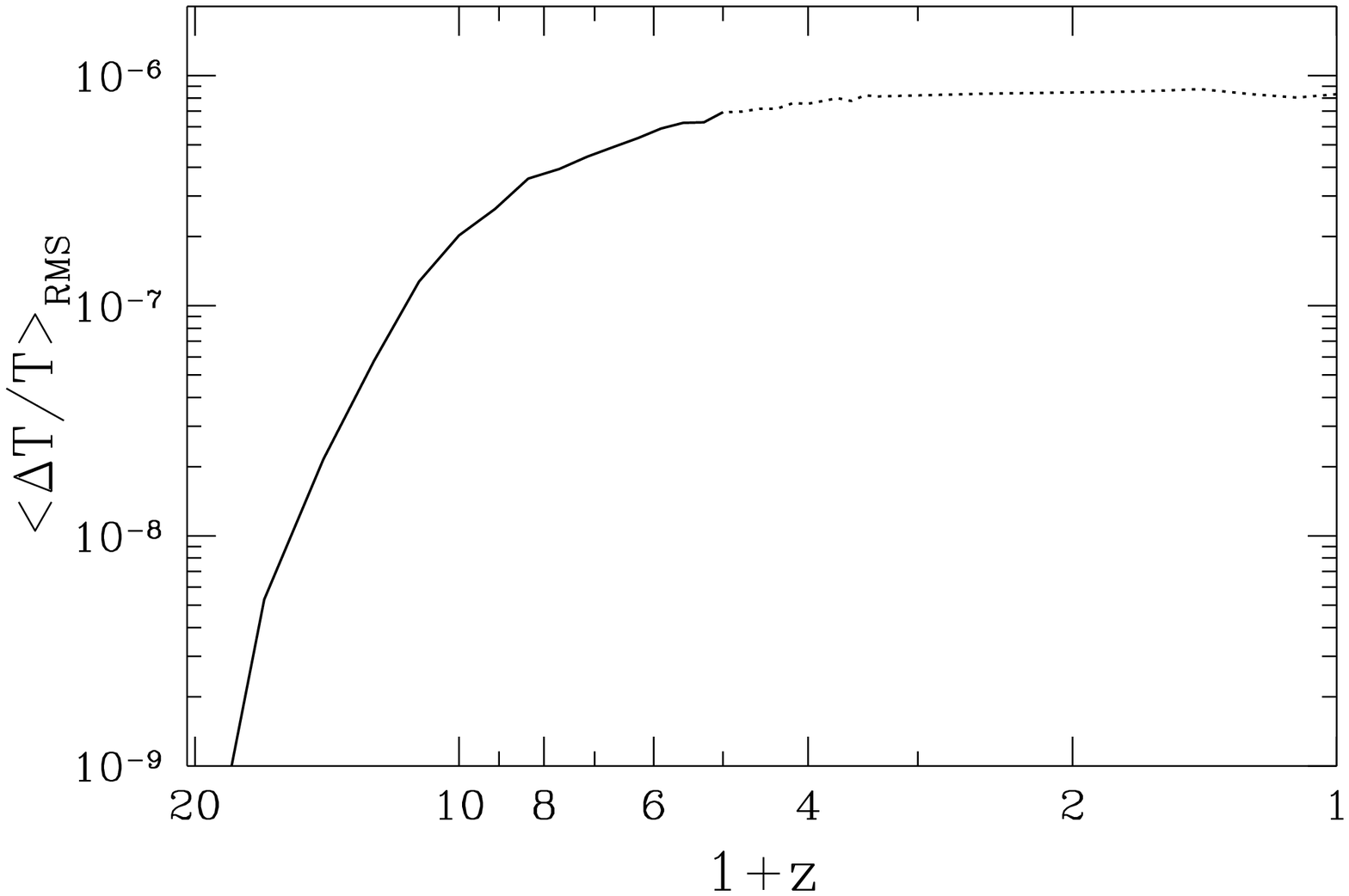}
\insertfigure{\figdir/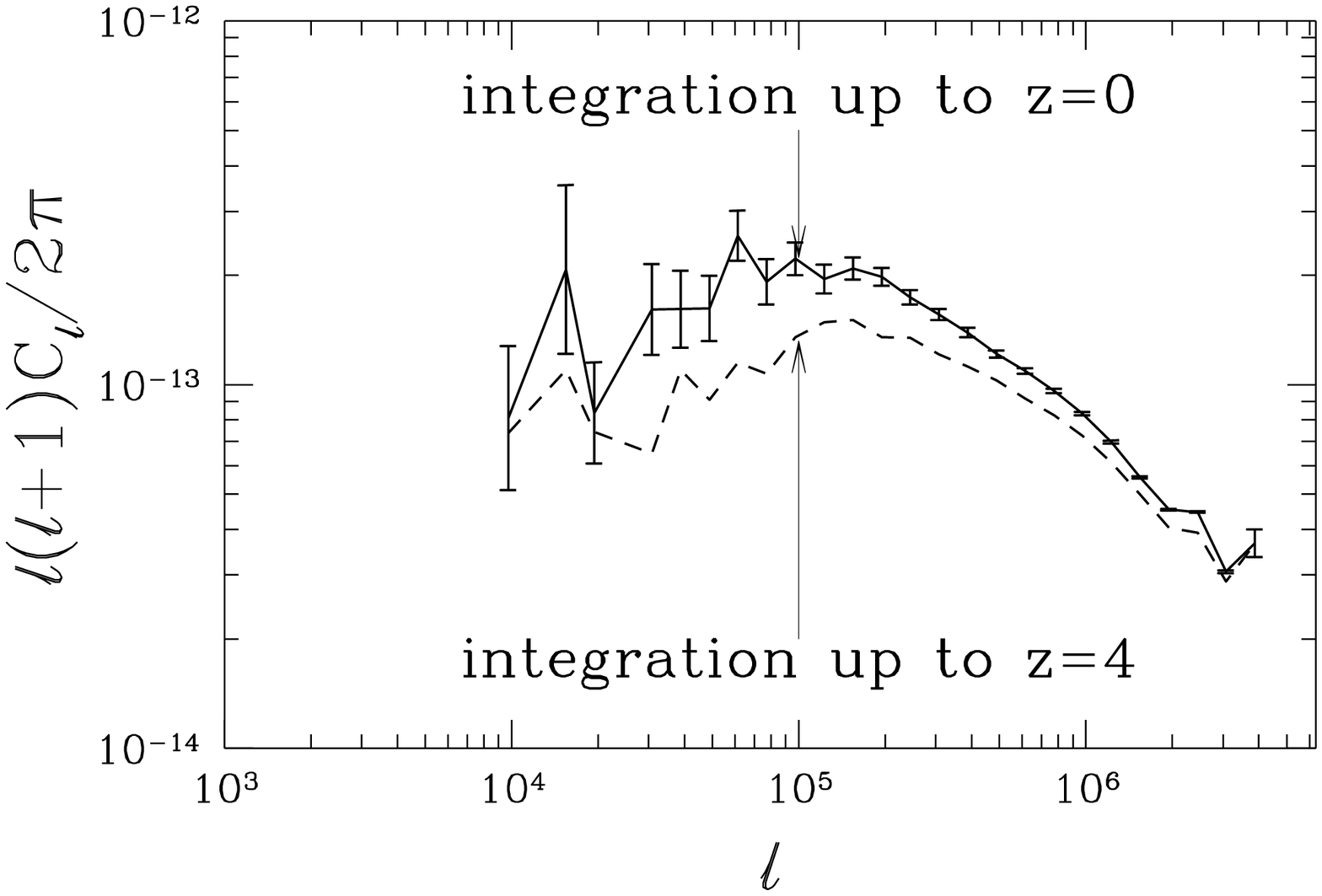}
\caption{\label{figNC}\capNC}
\end{figure}
} 
Finally, since our simulation is stopped at $z=4$, we need to estimate
the contribution from the lower redshifts. Since the universe is fully
ionized at $z=4$, this contribution is entirely due to a homogeneous
ionization fraction, $x_e=1$ (the NL Ostriker-Vishniac effect).  We can
calculate it by extrapolating from higher redshifts. Figure \ref{figNC}a
shows the rms CMB temperature anisotropy as a function of the final
redshift of calculation $z$ from our simulation (up to $z=4$) with the
solid line. One can see that even at $z=4$ the rms temperature
anisotropy continues to rise (despite the fact that the universe is
fully ionized by then), and thus we miss a  portion of the
total signal. In order to estimate the missing signal, we extrapolate
the signal to $z<4$ assuming that the density and velocity structure
remains fixed in the comoving coordinates (in other words, taking the
output of the simulation at $z=4$ and adopting it for all lower
redshifts). Then we find that
$$
        \left.\Delta T\over T\right|_{\rm RMS}\!\!\!\!\!(z=0) = 1.25 
        \left.\Delta T\over T\right|_{\rm RMS}\!\!\!\!\!(z=4),
$$
i.e., we miss about 20\% of the signal. This number is likely to be an
overestimate, as the gas pressure effects will erase the structure on
spatial scales up to $500\dim{kpc}$ by $z=0$, which corresponds to
angular scales which are much larger than the ones dominating the
signal.  Nevertheless, since we cannot continue the simulation beyond
$z=4$, we use the extrapolated power spectrum as our final
result. The two spectra, the extrapolated one and the one directly
computed from the simulation, are shown in Fig.\ \ref{figNC}b. The
difference between the two is larger than, albeit comparable to, the
statistical error due to the finite box size of the simulation, and is
the largest uncertainty of our calculation on small angular scales
(must less than the size of the box). On the angular scales comparable to the
box size, an additional uncertainty is introduced by the missing large-scale 
power.

\subsection{Perturbations Outside of the Simulation Box}

Since the size of the simulation box is fixed to $4h^{-1}\dim{Mpc}$ in
comoving units, we are unable to calculate directly anisotropies due to
perturbations on larger scales. However, we can include them
analytically using the linear theory. 
Since we are
not able to follow the ionization state of the gas on these scales, we
assume that in the linear regime the ionization fraction is spatially
homogeneous.  This is not a bad approximation after all, since the size
of a typical $\HII$ region is smaller than the simulation box size, and
thus on larger scales we average over a sufficiently large volume of the
universe.  The latter implies that in the linear regime we can consider
only the Ostriker-Vishniac effect -- the NLOV effect is the same as the
LOV effect. As we argue below, it is the dominant contribution to the
total anisotropies on all angular scales.

To calculate the pure LOV effect (linear fluctuations and homogeneous
reionization), we use the approach described in Jaffe \& Kamionkowski
(1998). We use a small-angle Fourier-space version of Limber's equation
which decouples the line-of-sight projection from the three-dimensional
power spectra. As discussed above, we also assume the linear
relationship between density and velocity, uniform reionization, and
Gaussian statistics. This reduces the calculation to a multiple integral
over a product of the linear power spectrum of density perturbations
with itself.

To check our methods against this analytic calculation, we perform an
additional ``simulation'' of the LOV effect in which we assume that the
density and velocity fields evolve only according to linear theory, and
we assign the ionization fraction at a given redshift uniformly over the
computational box with the value given by the volume average ionization
fraction in the full numerical simulation. We then make an image and
calculate the CMB power spectrum in precisely the same fashion as in the
full simulation. Thus, the spectrum of the anisotropies from a such
``linear'' simulation includes all the finite-box effects present in the
full numerical simulation.

\def\capLT{
The spectrum of the CMB temperature anisotropy as calculated
  in linear theory ({\it short-dashed line\/}), from the full numerical
  simulation ({\it bold solid line\/}), and from the ``linear''
  simulation ({\it dotted line\/}).  Also shown with the long-dashed
  lines the nonlinear power spectra (for two box sizes) of kinetic SZ
  effect on large scales from Springel et al.\ (2000).
}  
\placefig{
\begin{figure}
\insertfigure{\figdir/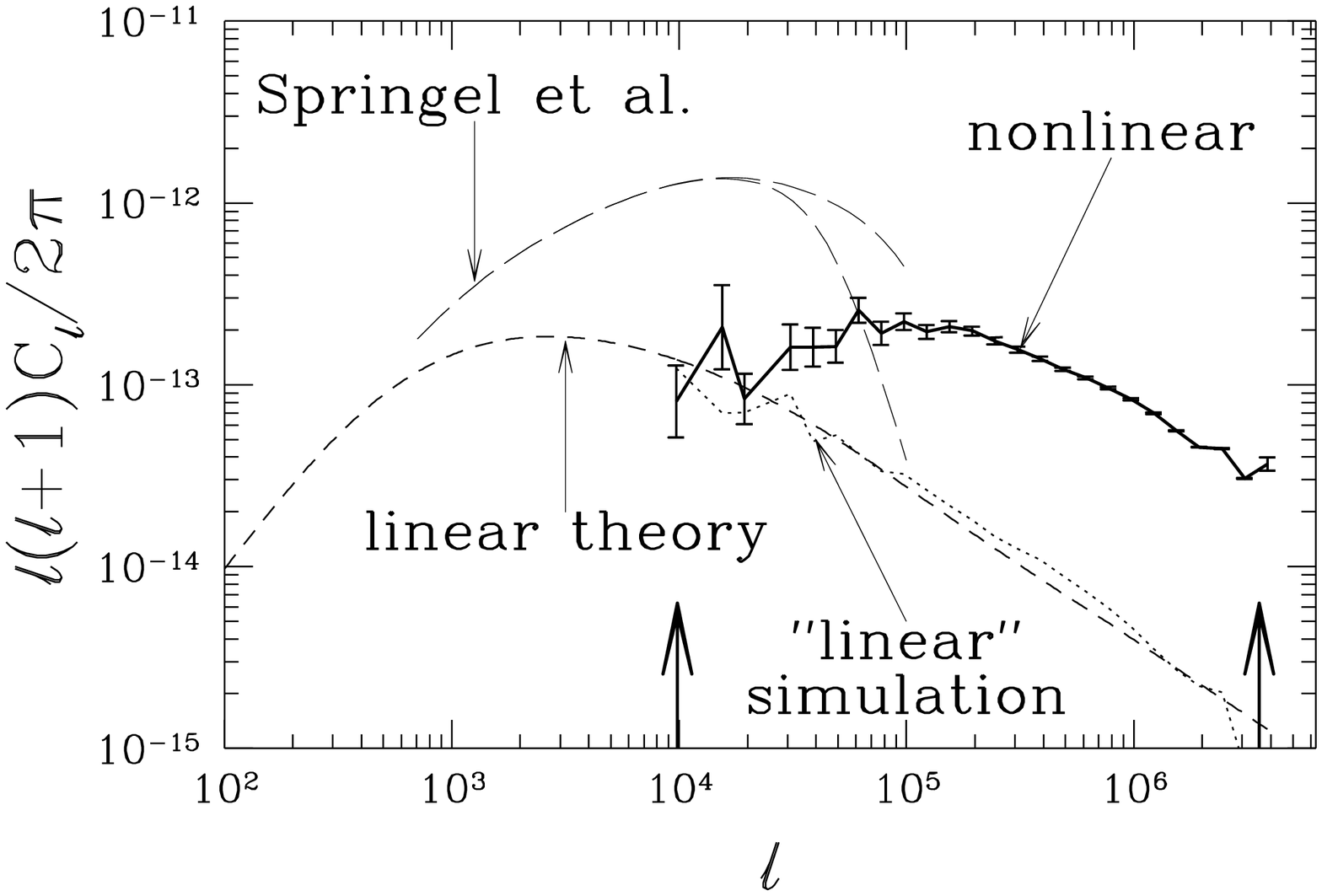}
\caption{\label{figLT}\capLT}
\end{figure}
}
\def\capIM{
Secondary temperature anisotropies in a 
$2.2^\prime\times2.2^\prime$ patch on the sky as generated during
reionization ({\it left panel\/}) and 
with randomized phases (i.e., a realization of a Gaussian with the same 
power spectrum, {\it right panel\/}). The color bars show the color-scale
correspondence, where the anisotropies are measured in $\mu K$.
}
\placefig{
\begin{figure}
\insertfigure{\figdir/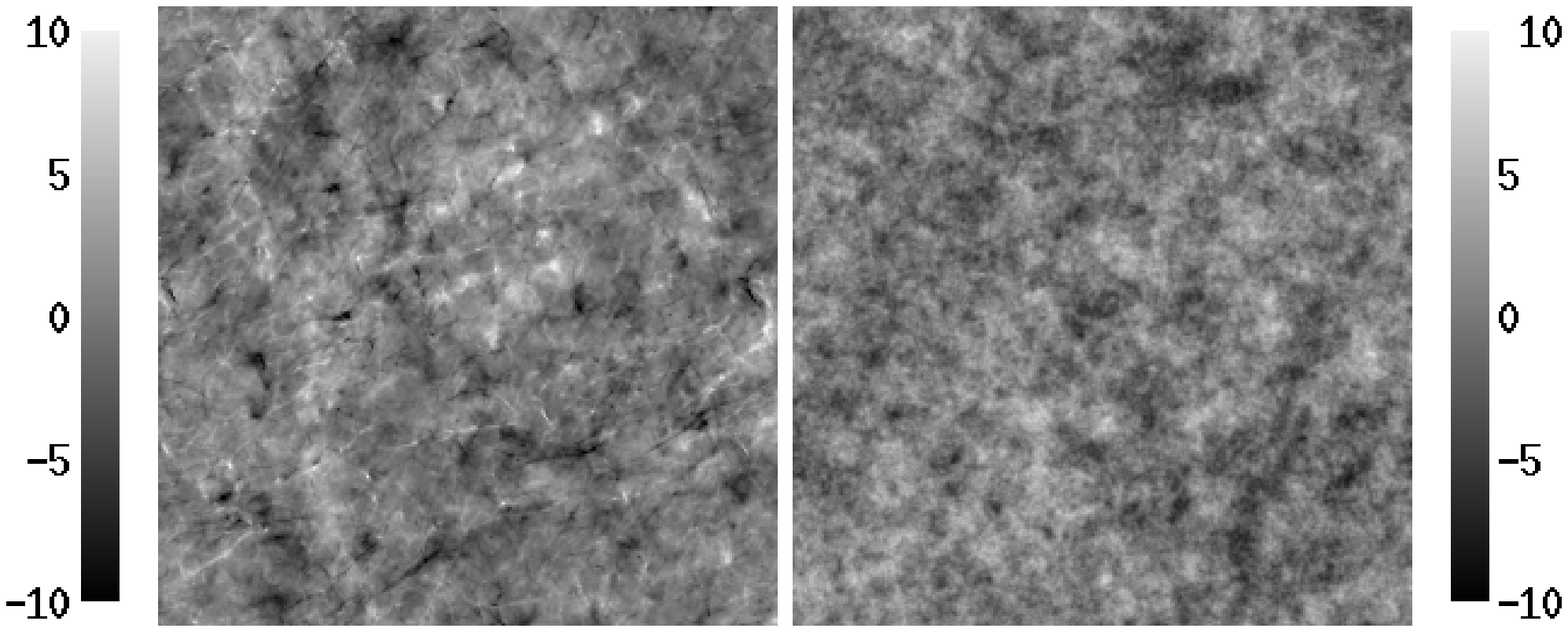}
\caption{\label{figIM}\capIM}
\end{figure}
}
The full nonlinear power spectrum is shown with the bold solid line
in Fig.\ \ref{figLT}. In addition, we show the output from our linear
``simulation'' along with the analytic OV calculation, which match over
all scales computed. We emphasize that this agreement requires the
imposition of appropriate velocities on scales larger than the 
simulation box, as
discussed above. 
We also note that our full nonlinear prediction (bold solid line)
matches the linear theory
on the angular scale corresponding to the box size. This is of course
is an artifact of our procedure (and the limited size of our simulation
box), which does not include nonlinear perturbations outside the simulation
box. In reality, there exists additional power on scales which we can not
resolve, simply because the nonlinear scale at $z=0$ is about twice
larger than our whole box.
To account for this power, we show the nonlinear power spectrum 
computed by
Springel et al.\ (2000). They assumed precisely the same cosmological
model and the redshift of reionization. In addition, 
their mass resolution almost exactly
corresponds to 1/8 of the total mass of our simulation box
(corresponding to 1/2 the box size - the largest scale which can be
resolved in our simulation). Thus, Springel et al.\ (2000) begin
precisely at the scale where our calculation ends, and we will use their
results to compliment ours below. (Conversely, this means that we
unfortunately cannot actually check the calculations against one
another.)

\section{Results}

\subsection{Maps and non-Gaussianity}

It is customary to characterize the CMB temperature anisotropies on the sky
by the power spectrum $C_\ell$.
The power spectrum is sufficient to fully
describe properties of a physical quantity only if this quantity is
Gaussian distributed. For the primary temperature anisotropies this
is indeed the case, since they are generated in the linear regime.
Reionization however is a non-linear process, and 
secondary anisotropies we consider in this paper are not necessarily
Gaussian. 

This is illustrated in Figure \ref{figIM}, which
shows the actual maps of the CMB anisotropies on
the square patch on the sky in the simulation.
The left panel shows the actual anisotropies on
2.2 arcmin patch on the sky 
(the angular size of our computational box at $z=8$, just before the full
overlap of \HII\ regions). The right panel, however,
shows anisotropies with the same power spectrum but 
distributed as a Gaussian. We notice that 
the actual map is highly non-Gaussian: 
it is much more structured than the Gaussian version
and contains higher fluctuations than would be expected in a Gaussian
field.

\def\capPD{One-point probability distribution of the temperature anisotropies
({\it solid line\/}), the one-point
probability distribution of spherical multipole amplitudes $a_{\ell m}$
({\it dashed line\/}),
and a Gaussian distribution ({\it dotted line\/}), all
normalized to unit dispersion.}
\placefig{
\begin{figure}
\insertfigure{\figdir/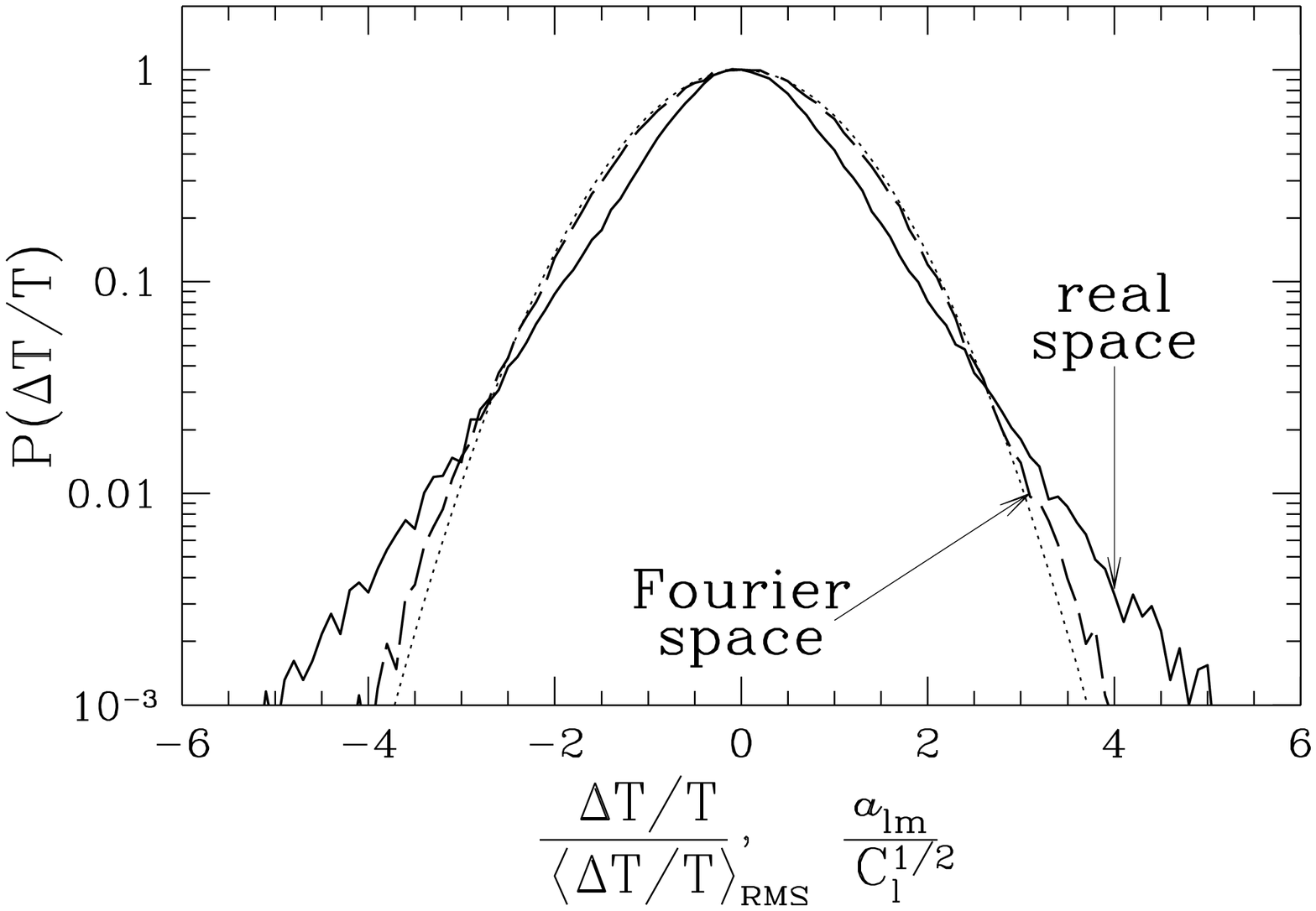}
\caption{\label{figPD}\capPD}
\end{figure}
}
\def\capCC{
The electron density correlation function $C_{ee}(\eta,\eta+\Delta\eta,\theta)$
as a function of angular size $\theta$ (measured in units of comoving
distance $R_\theta$ at this redshift) and $\Delta\eta$
(also measured in units of comoving distance $R_\eta$) at three redshifts
(values of $\eta$): 
({\it a\/}) $z=10$,
({\it b\/}) $z=7$, and
({\it c\/}) $z=5$.
Subsequently heavier shaded contours show increase in the correlation function
by a factor of 10. The outmost contour corresponds to $C_{ee}=1$.
}
\placefig{
\begin{figure}
\epsscale{0.90}
\insertthreefigures{\figdir/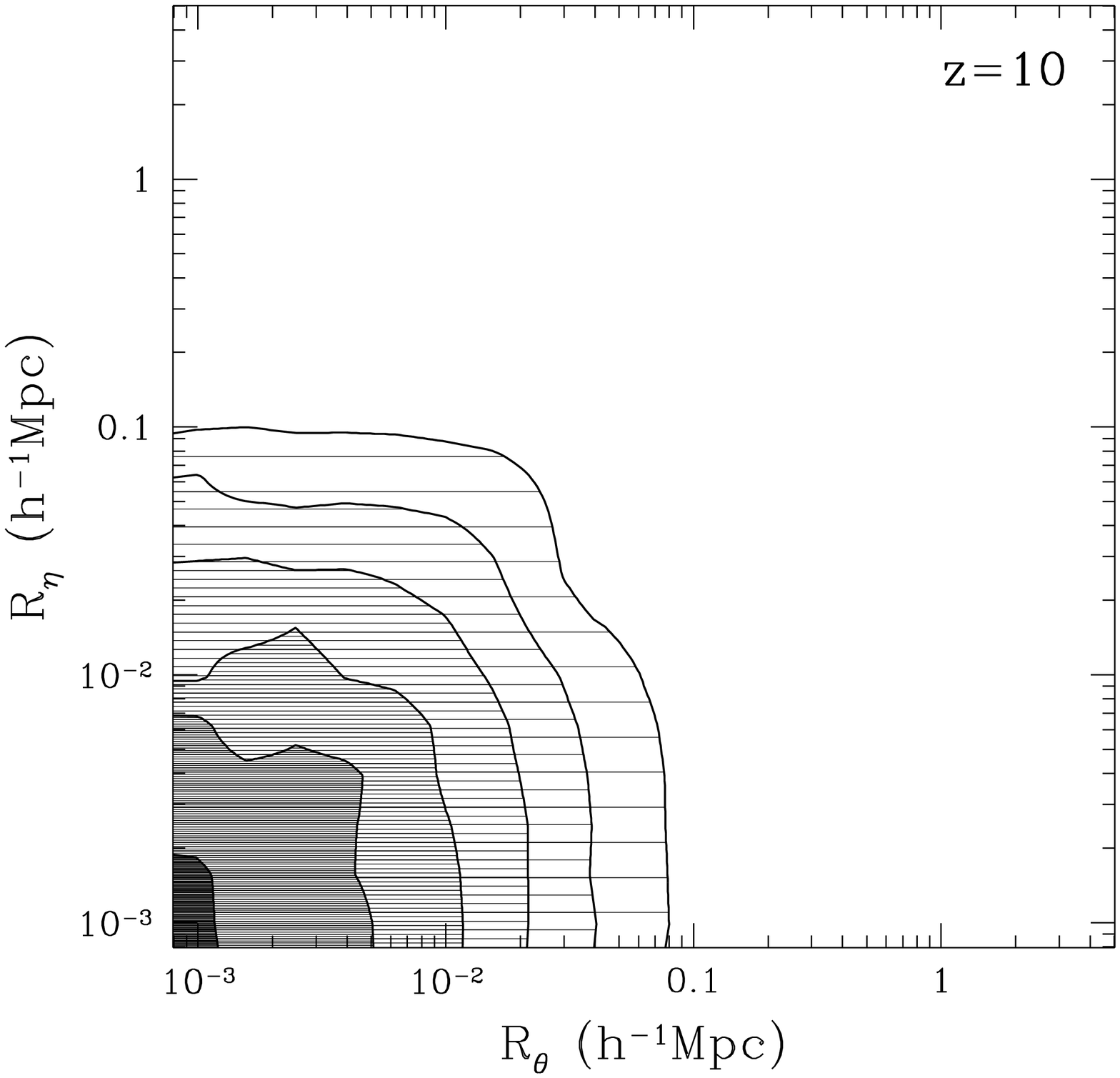}{\figdir/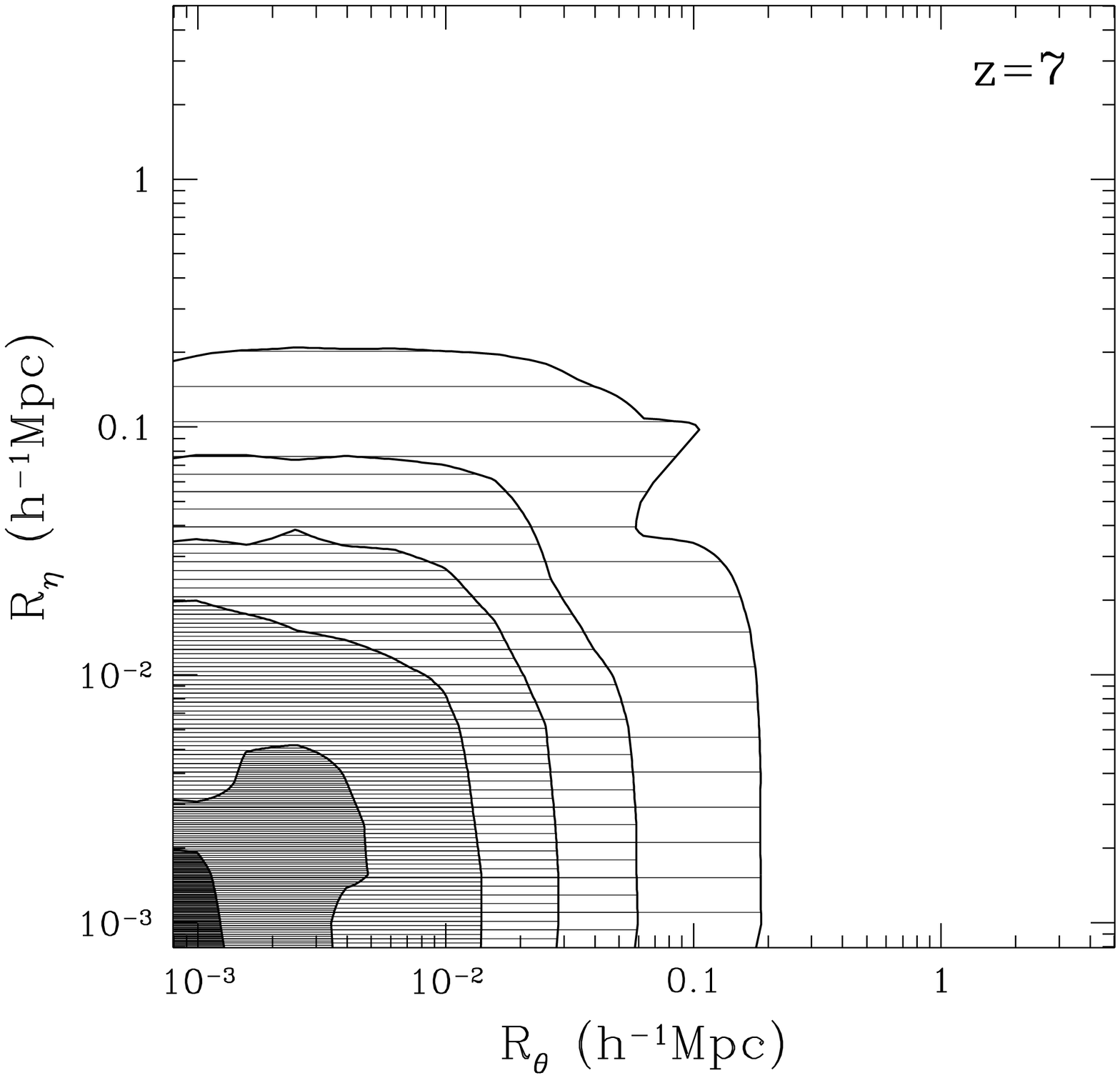}{\figdir/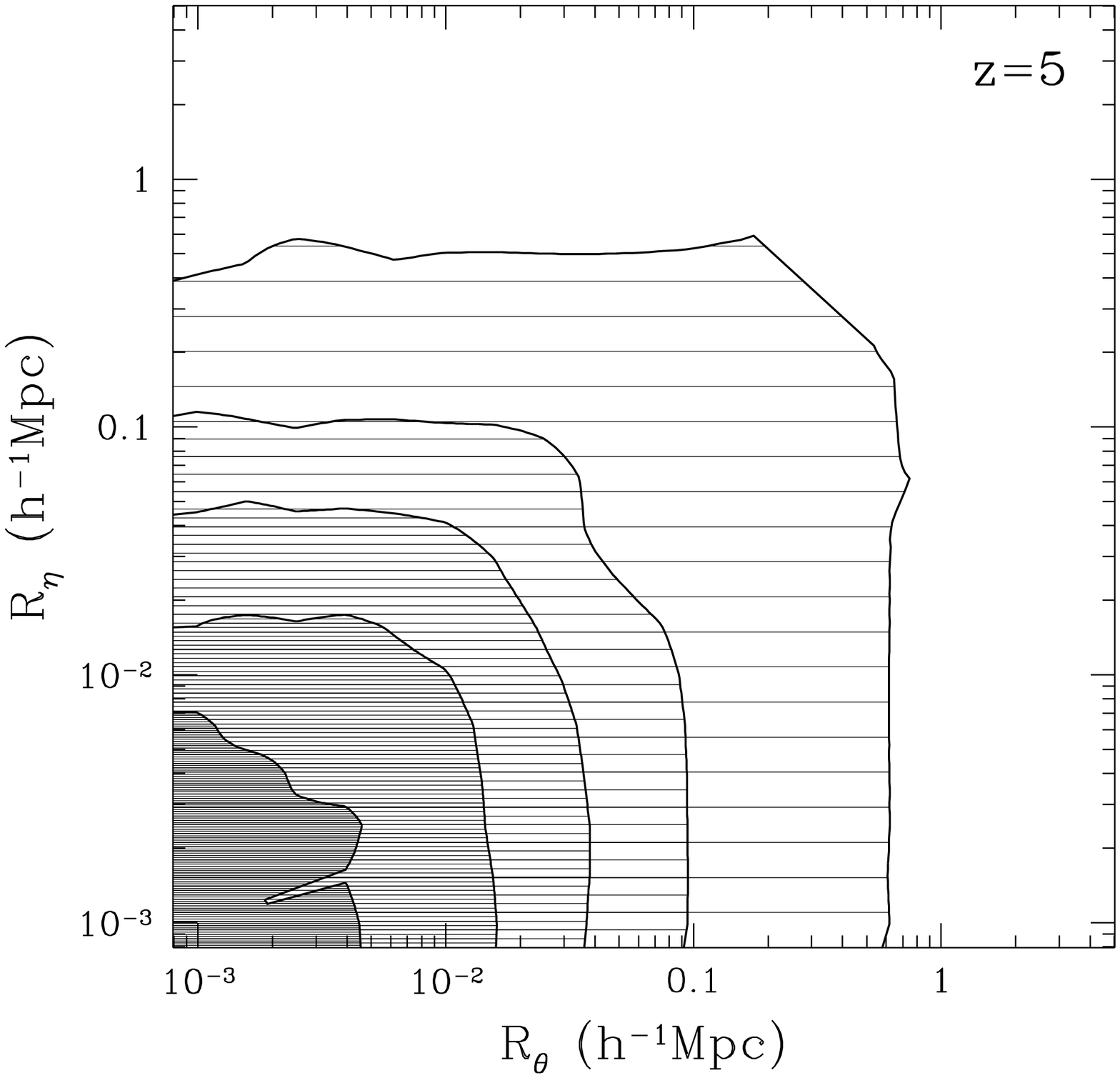}
\caption{\label{figCC}\capCC}
\end{figure}
}
To illustrate this further, we show in 
Figure \ref{figPD} the one-point distribution of the actual
temperature fluctuations, and a Gaussian distribution for comparison.
We notice that the temperature anisotropies are non-Gaussian, and
thus more information than simply $C_\ell$
can be extracted from the actual distribution on the sky. In addition
to temperature anisotropies on the sky, we also show in Fig.\ \ref{figPD} the
one-point distribution function of spherical multipole 
amplitudes $a_{\ell m}$, which
are defined in the usual fashion by expansion over spherical harmonics,
$$
        {\Delta T\over T}(\theta,\phi) = 
        \sum_{\ell,m} a_{\ell m} Y^m_\ell(\theta,\phi).
$$
The one-point distribution of the amplitudes is actually quite similar
to a Gaussian, which indicates that it is indeed the the phase
correlations between different multipoles what drive the non-Gaussian
patterns. In order to characterize this non-Gaussianity, one would have
to measure the higher-order moments (bispectrum, trispectrum, etc.) at
non-zero lags, or other quantities such as the Minkowski functionals,
but that is outside the scope of this paper.

\subsection{Correlation Functions}

We now investigate the behavior of the correlation functions. In particular,
the electron density correlation function $C_{ee}(\eta_1,\eta_2,\theta)$ is
of highest interest. We show in Figure \ref{figCC} the electron density
correlation function at three redshifts: before the overlap of \HII\ regions
at $z=10$, during the overlap at $z=7$, and after the overlap at $z=5$. One
can notice that there is no drastic change in the shape of the correlation 
function despite the qualitative change in the structure of the ionized
regions; only the correlation length increases with time. This indicates
that the signal is dominated by the ionized high density regions at all
times, and is not sensitive to the ionization state of the low density 
intergalactic medium. We will elaborate on this below.

\def\capCO{
The electron density correlation function $C_{ee}(\eta,\eta,\theta)$
({\it solid line\/}) and the velocity correlation function
$C_{vv}(\eta,\eta,\theta)$ ({\it dashed line\/}) at $z=7$. The dotted
line shows the quantity $C_{ev}^2/C_{vv}$.
}
\placefig{
\begin{figure}
\insertfigure{\figdir/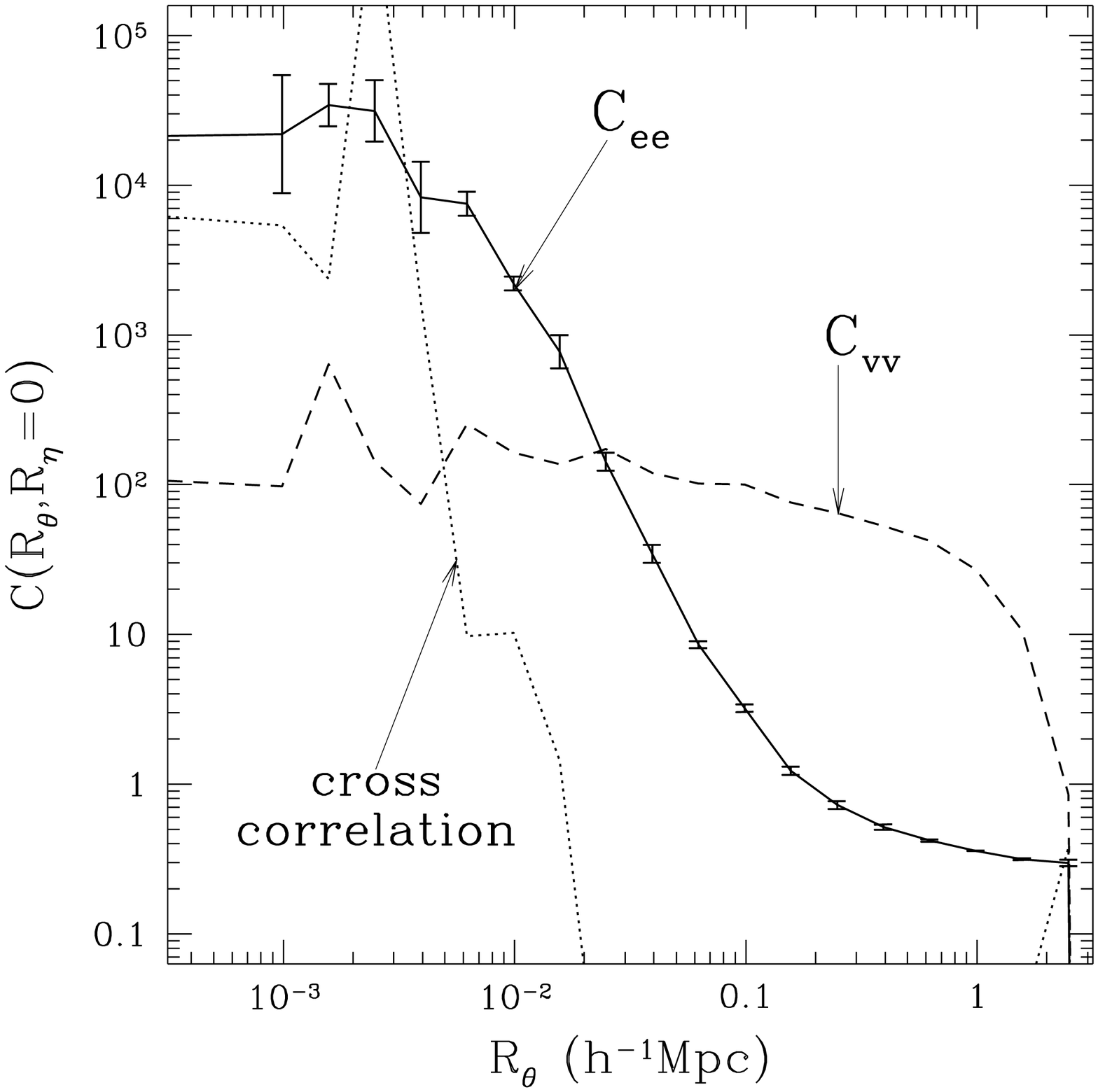}
\caption{\label{figCO}\capCO}
\end{figure}
} Figure \ref{figCO} shows both the electron density correlation
function $C_{ee}$ and the velocity correlation function $C_{vv}$ as a
function of angle (measured as comoving distance at this redshift) at
$z=7$. We also show with the dotted line the quantity $C_{ev}^2/C_{vv}$,
which is much smaller than $C_{ee}$ if the cross-correlation between the
electron density and velocity is not significant, and is equal to
$C_{ee}$ is they are completely correlated.  One can see that the
cross-correlation only becomes significant on very small scales, about
0.1 arcsec ($\ell\sim6\times10^6$), 
and thus the usual assumption that velocity and electron
density are uncorrelated is highly accurate on all scales of interest.

\subsection{Kinetic vs Thermal SZ Effect}

\def\capTT{
Comparison of the power spectrum of the secondary CMB anisotropies in the
computational box 
generated by the kinetic SZ effect (eq.\ [\ref{eq:startingpoint}], 
{\it solid line\/}) and the thermal SZ effect
(eq.\ [\ref{thermalsz}], {\it dashed line\/}).
}
\placefig{
\begin{figure}
\insertfigure{\figdir/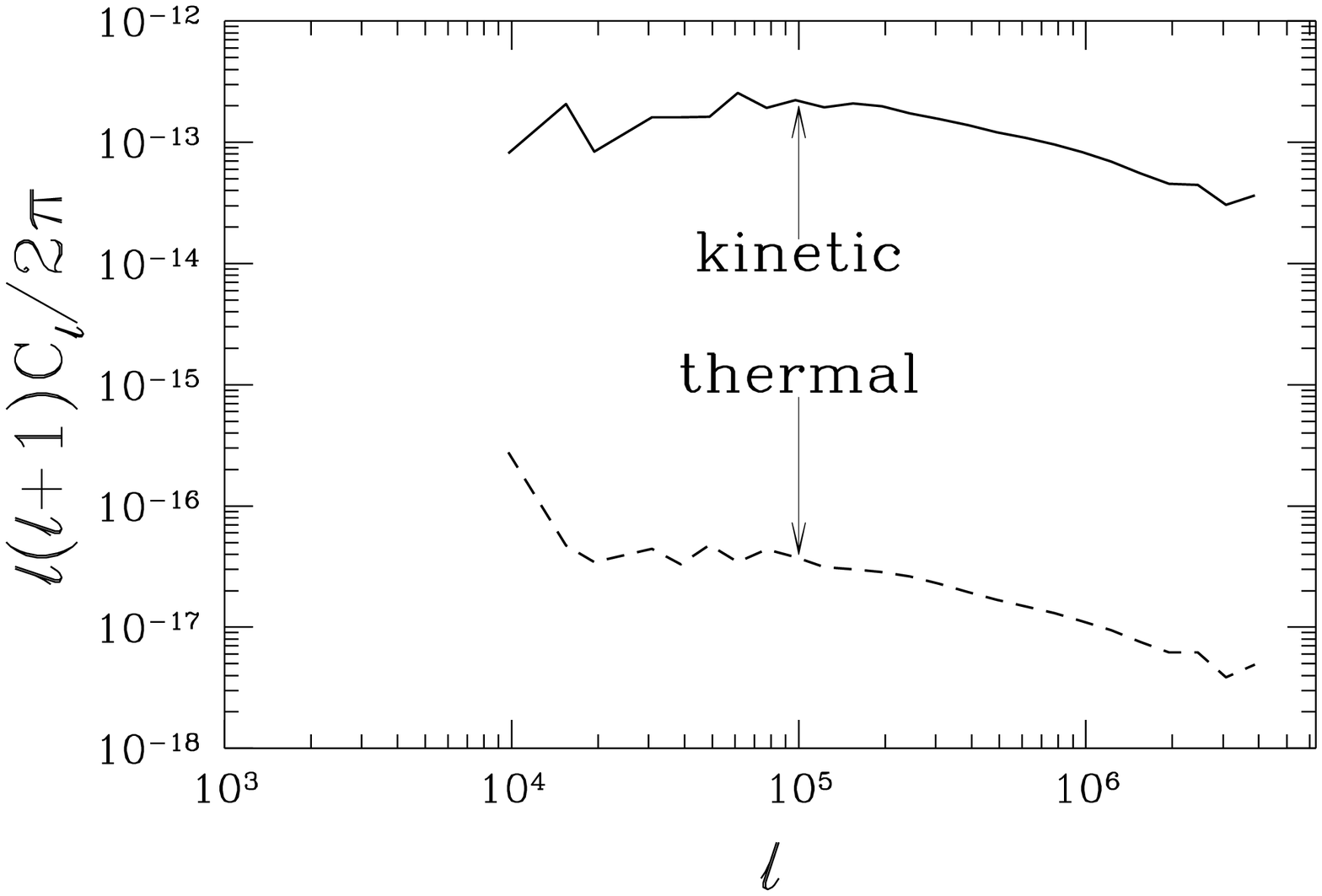}
\caption{\label{figTT}\capTT}
\end{figure}
}
It is interesting to compare the kinetic SZ effect given by equation
(\ref{eq:startingpoint}) with the thermal SZ effect, given by
equation (\ref{thermalsz}). One can expect a priori that the kinetic SZ
effect, generated in the gas moving with some $300\dim{km}/\dim{s}$ is
about 1000 times more important (in terms of the power spectrum) than the 
thermal SZ effect, generated due to thermal motions in the $10^4\dim{K}$
hot gas. Indeed, this is demonstrated in Figure \ref{figTT}.
Thus, for all practical
purposes, the kinetic effect is the only one that needs to be taken
into account on these scales. Of course, for clusters of galaxies, the
situation is reversed and the temperature fluctuation induced by even
the several-hundred km/s bulk motion of the cluster is dwarfed by the
thermal effect, simply because the cluster gas temperature is some four
orders of magnitude higher than the IGM temperature, whereas the bulk 
gas velocity in a cluster is only about a few times that of the IGM. We
return to the comparison between the two effects below.

\subsection{The Ostriker-Vishniac Effect versus Patchy Reionization}

\def\capCV{
The electron density correlation function $C_{ee}(\eta,\eta,\theta)$
for the patchy reionization effect
($C_{ee}^{PR}$, {\it solid line\/}) and for the NL Ostriker-Vishniac
effect ($C_{ee}^{OV}$, {\it bold dashed line\/}) at $z=9$ (at other
redshifts the two correlation functions look quite similar except for the
change of the characteristic scale). We also show with the dotted
line the quantity $\left(C_{ee}^{OV-PR}\right)^2/C_{ee}^{OV}$, which
coincides with $C_{ee}^{PR}$ if cross-correlations are important. The thin
dotted line shows the NL Ostriker-Vishniac correlation function
$C_{ee}^{OV}$ rescaled by the quantity 
$\left[(1-\bar{x}_e)/\bar{x}_e\right]^2$.
}
\placefig{
\begin{figure}
\insertfigure{\figdir/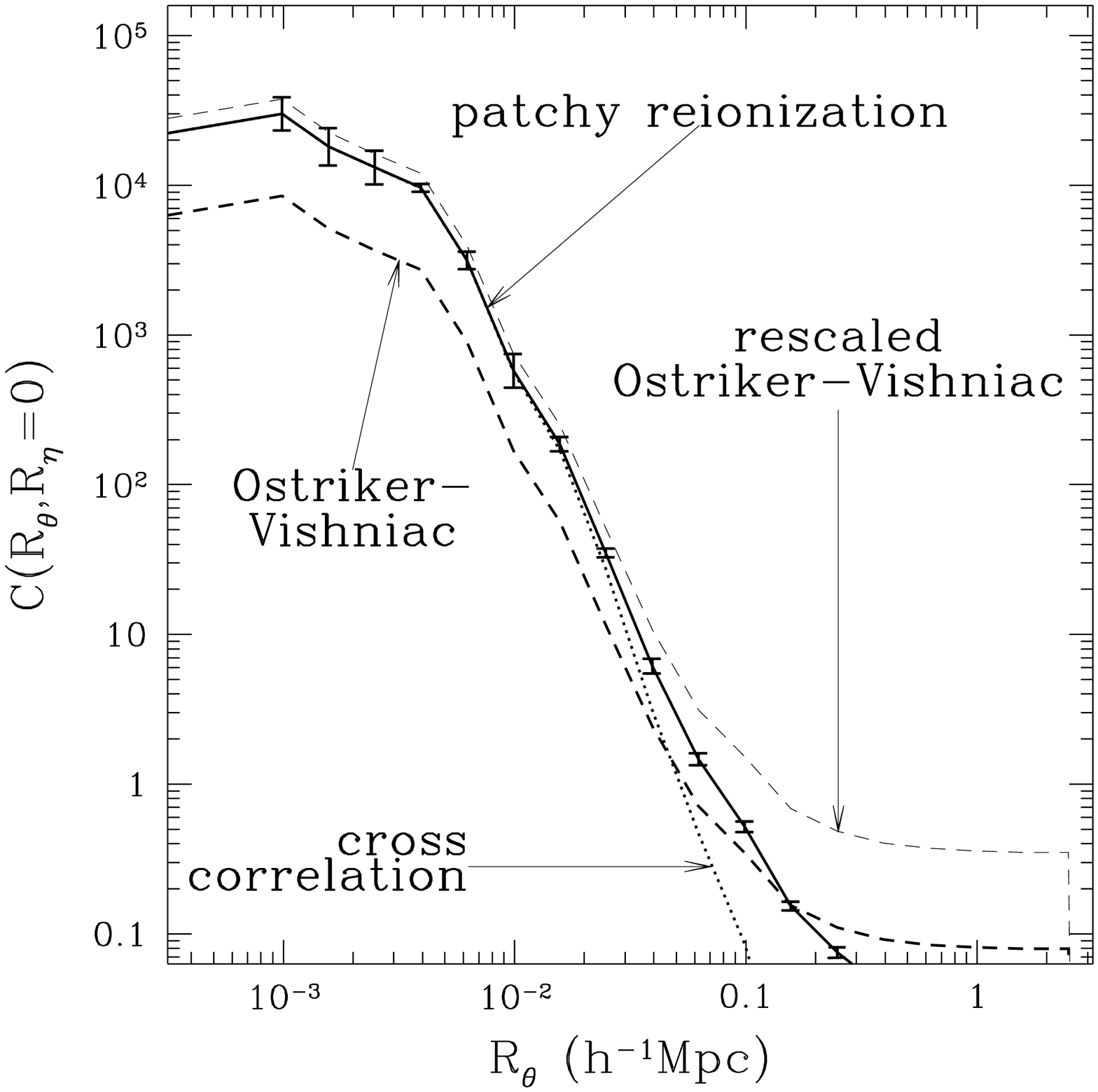}
\caption{\label{figCV}\capCV}
\end{figure}
}
We now focus on a comparison between the NL Ostriker-Vishniac effect and 
patchy reionization. Figure \ref{figCV} shows the comparison between
the two correlation functions $C_{ee}^{PR}$ and $C_{ee}^{OV}$, which are
defined by equations (\ref{eq:ovprcf}). The dotted
line, marking the cross-correlation term in comparison with $C_{ee}^{PR}$,
only falls significantly below $C_{ee}^{PR}$ at large radii ($R>0.1h^{-1}
\dim{Mpc}$, where the inhomogeneous distribution of the ionized fraction due to
expanding $\HII$ regions make the NL Ostriker-Vishniac effect and 
patchy reionization effect uncorrelated. However, as the thin dashed line 
shows, in the high density regions (i.e., on small scales) the effect of
patchy reionization is almost equivalent to having $x_e=1$, i.e., to having
almost all high density regions ionized. This implies that the bulk of the
signal comes from high density regions and correlations between them.

In order to investigate the relationship between the two effects
further, we have constructed a simulation which contains only the
``Ostriker-Vishniac'' effect by assigning a uniform ionization fraction
(equal to the volume average at a given redshift) to the outputs of our
full numerical simulation, and performing line-of-sight integration as
described above. Thus, this ``NLOV simulation'' has the density and
velocity structure of our full numerical simulation, and has the same
evolution of the volume averaged ionization fraction, but has uniform
ionization at all densities and spatial locations. In particular,
the Patchy Reionization correlation function $C_{ee}^{PR}$ is equal
to zero
in that simulation, and its NL Ostriker-Vishniac
correlation function ($C_{ee}^{OV}$) is the same as in the full simulation.
Thus, the NLOV simulation allows us to construct an image of the CMB
anisotropies which correspond to just the $C_{ee}^{OV}$ correlation function,
whereas the full simulations gives us anisotropies corresponding to the
sum of the NL Ostriker-Vishniac and Patchy Reionization effects.

\def\capRV{
The rms CMB temperature anisotropy as a function of the final redshift of
integration for 
the NL Ostriker-Vishniac effect ({\it dashed line\/}),
the patchy reionization effect ({\it dotted line\/}), and for
the total effect (the sum of the two, {\it solid line\/}).
}
\placefig{
\begin{figure}
\insertfigure{\figdir/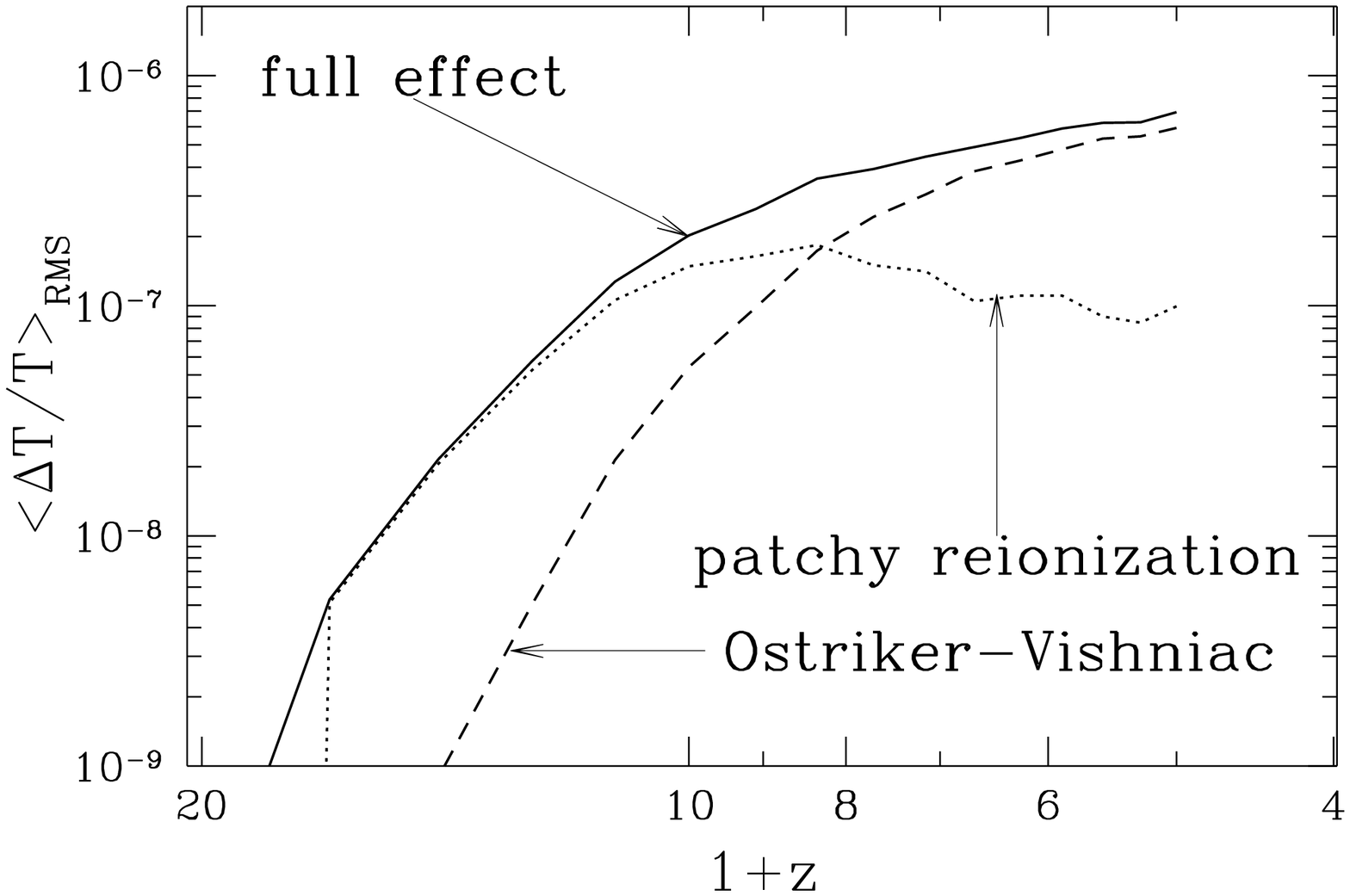}
\caption{\label{figRV}\capRV}
\vspace*{3in}
\end{figure}
}
Figure \ref{figRV} now shows
the rms temperature anisotropy on the sky (for the $256^2$ pixelization)
for the total effect, and separately for the NL Ostriker-Vishniac effect
(as computed from the ``NLOV simulation'')
and for the patchy reionization effect (computed as the difference 
between the two). Note, that by $z=4$ almost all the signal 
comes from the NL Ostriker-Vishniac effect, and the contribution of the
patchy reionization actually slightly 
declines at $z<7$, after the overlap of
$\HII$ regions. This is due to the fact that after the overlap the
topology of the ionized phase flips over: reionization starts with 
isolated ionized regions expanding into the neutral medium, and ends
with the isolated neutral regions being ionized from outside inward, which
produces an anti-correlation between the ionized and neutral phases and leads
to suppression of anisotropies due to patchy reionization.

\def\capCL{ The solid line shows the total kinetic SZ power spectrum of the 
 secondary CMB anisotropies, computed by combining our full nonlinear
 results (the solid line from Fig.\ \ref{figLT}) for $\ell>10^4$ with
 the  results of 
 Springel et al.\ (2000) for $10^2<\ell<3\times10^4$ and linear theory 
for $\ell<10^2$), whereas the long-dashed line shows the same calculation
for the thermal SZ effect. For both the kinetic and thermal effects, the bold
  portions of the lines show the regions of the spectrum reliably
  calculated in this work (large $\ell$ tail), in Springel et al.\ 
  (2000), or in linear theory, and the thin line shows the regions of
  $\ell$ where existing predictions are unreliable (and thus the curves
  we show are somewhat arbitrary).
  Two dotted lines show the result for the patchy reionization,
  calculated from our simulation as the difference between the full
  simulation and the ``NLOV simulation'' ({\it bold dotted
    line\/}), and an approximate prediction from Gruzinov \& Hu (1998)
  ({\it thin dotted line\/}).  Also shown is the spectrum of primary
  anisotropies for the assumed cosmological model ({\it thin solid
    line\/}).
} \placefig{
\begin{figure}
\insertfigure{\figdir/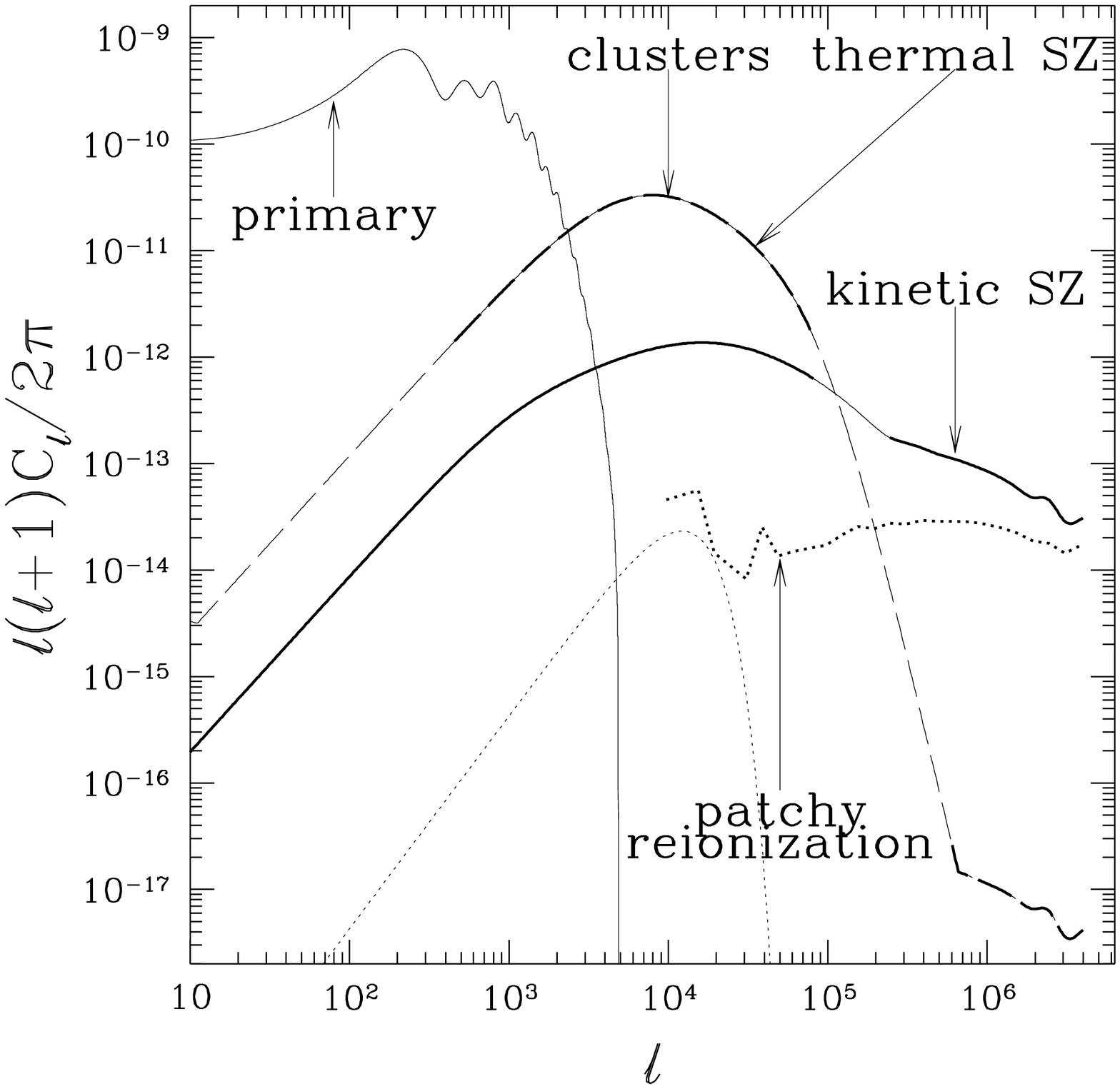}
\caption{\label{figCL}\capCL}
\end{figure}
}
Figure \ref{figCL} presents now the main result of this paper: the
power spectrum of the secondary CMB anisotropies from the kinetic SZ
effect, from the thermal SZ
effect, and from the patchy reionization effect. For the former two, we
combined our results with the calculations by Springel et al.\ (2000).
In the range $3\times10^4\leq\ell\leq2\times10^5$ our calculation
for the kinetic SZ effect disagrees
with the extrapolation from the Springel at al.\ (2000), and for the
thermal SZ effect the range of disagreement is even larger.
 Since our results
can suffer from the missing large-scale power at low redshift, and 
small scale extrapolation  of Springel et al.\ (2000) is likely to be
severely affected by limited numerical resolution, it is impossible to
give a reliable prediction for the CMB power spectra in this range of
angular scales, and we emphasize this fact by using the bold lines to
show the calculations where they are reliable, and by using the
thin lines in the range of scales where consider calculations to be
unreliable.

We also show the prediction from an analytical
calculation of the patchy reionization effect by Gruzinov \& Hu (1998)
for our cosmological model. Note that this prediction is a factor of
$10^2$ lower than the one given in Gruzinov \& Hu (1998) for the
standard CDM model, because they adopted the value for the rms velocity
at $z=0$ of $1200\dim{km}/\dim{s}$, whereas in our model this number is
about 3 times smaller, in agreement with observations (Baker, Davis, \&
Lin 2000). There is an additional reduction in the amplitude of the
effect by a factor of 10 due to the redshift of reionization being 7 in
our simulation compared to an adopted value of 30 in Gruzinov \& Hu
(1998).  Including all the relevant factors, we notice that Gruzinov \&
Hu (1998) estimate gives roughly the right prediction for the patchy
reionization effect on large scales. Of course, since the analytical
estimate of Gruzinov \& Hu (1998) does not include the density structure
on small scales, they are unable to reproduce the anisotropies on small
angular scales.

\subsection{Dependence on the Redshift of Reionization}

Up to know we assumed a fixed value for the redshift of reionization, since
we only used one simulation. In order to investigate the dependence of the
power spectrum on the redshift of reionization properly, we would need 
to run several simulation with different reionization histories, which is
not realistic at the current moment due to required computational resources.
However, in order to illustrate the dependence of anisotropies
on the redshift of reionization, and using the fact that at $z_{\rm REI}=7$ 
the NL Ostriker-Vishniac dominates the contribution to the
anisotropies due to Patchy Reionization, we have computed three additional
``NLOV'' simulations where we changed the redshift of reionization by
simply rescaling the volume averaged ionization fraction as a function of
redshift. In other words, if $n_{e,0}(a)$ is the electron density and
$\bar{x}_e(a)$ is the volume averaged ionization fraction 
as a function
of the scale factor for the
original ``NLOV'' simulation, for rescaled ``NLOV'' simulations we assumed
$$
        n_e(a) = {\bar{x}_e(8a_{\rm REI}a)\over \bar{x}_e(a)} n_{e,0}(a),
$$
where the factor 8 is simply $1/0.125=1/a_{\rm REI}=1+z_{\rm REI}$ 
for our simulation.

\def\capDZ{
The power spectra of the secondary CMB anisotropies computed in
the ``NLOV'' simulation with four different values for the redshift
of reionization: 
$z_{\rm REI}=5$ ({\it dotted line\/}),
$z_{\rm REI}=7$ ({\it short-dashed line\/}),
$z_{\rm REI}=11$ ({\it long-dashed line\/}), and
$z_{\rm REI}=19$ ({\it solid line\/}).
}
\placefig{
\begin{figure}
\insertfigure{\figdir/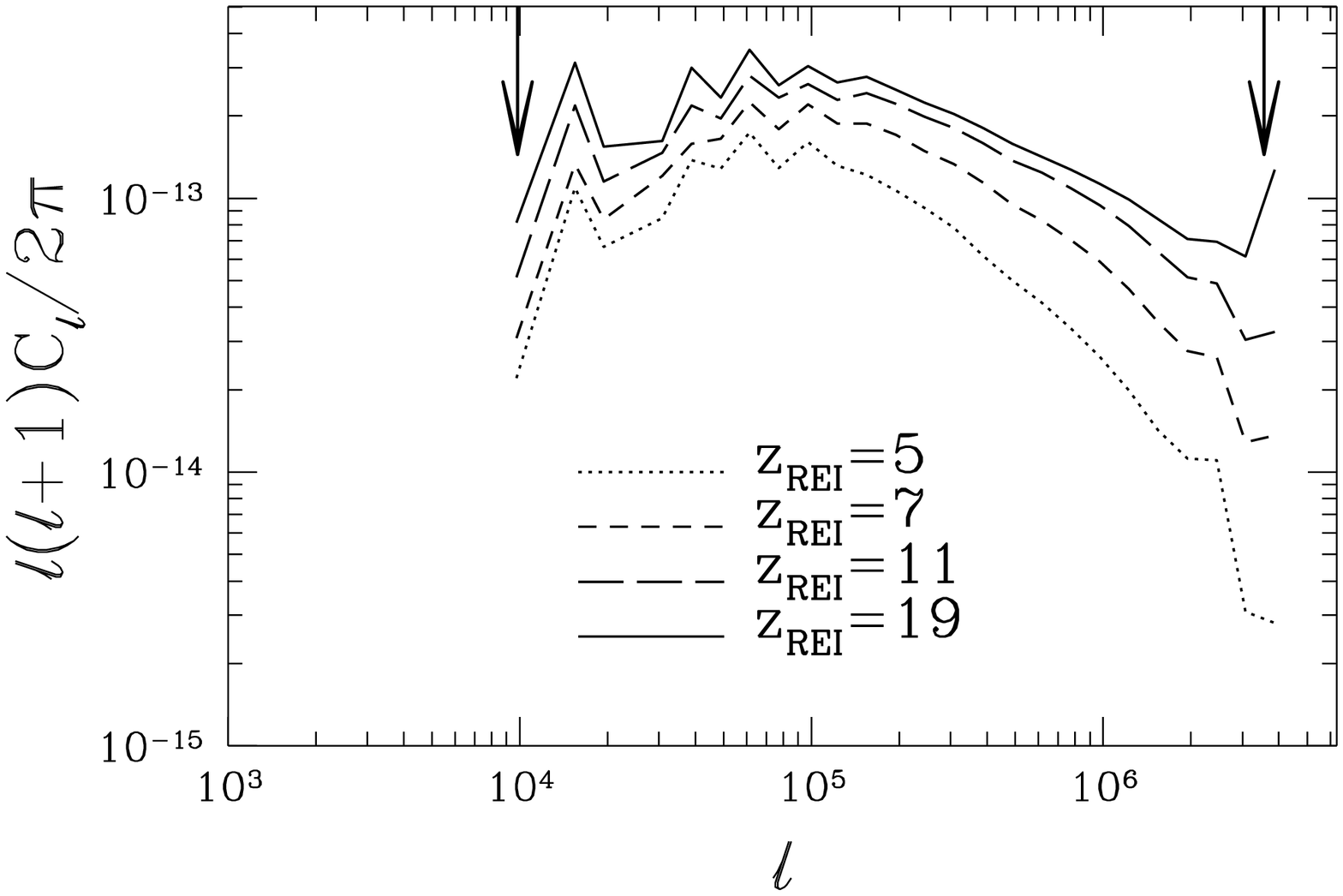}
\caption{\label{figDZ}\capDZ}
\vspace*{3in}
\end{figure}
}
Figure \ref{figDZ} shows the CMB power spectra for four different values for
the redshift of reionization in our rescaled ``NLOV'' simulations. We note
that in linear theory,
$\delta\sim a$ and $v\sim a^{1/2}$ at high redshift, and therefore the total 
anisotropy 
$$
        {\Delta T\over T} \sim \int^t a^{3/2} \bar{n}_b dt \sim \log(t)
$$
goes only logarithmically 
as a function of the moment of reionization (which is the upper limit in the
integral [\ref{eq:startingpoint}] for a model with sudden reionization). The 
dependence that we find in the nonlinear calculation is somewhat steeper
than that from the linear theory, but not by much.

We also note here that the analytical model of patchy reionization by
Gruzinov \& Hu (1998) predicts that the spectrum of the CMB anisotropies 
increases as $(1+z_{\rm REI})^{3/2} R_\HII$, where $R_\HII$ is 
the characteristic comoving 
size of $\HII$ regions, which is likely to go down as the redshift of
reionization increases. Thus, the total dependence on the redshift of 
reionization of the patchy reionization effect is somewhat slower than
$(1+z_{\rm REI})^{3/2}$ but perhaps not as slow as the nearly
logarithmic dependence of the NL Ostriker-Vishniac effect. Thus, if 
reionization occurred early, the role of patchy reionization (as compared to
the NL Ostriker-Vishniac effect) would be greater, albeit the current 
observational limits of $z_{\rm REI}\la30$ will not allow a regime when
patchy reionization becomes dominant.

\section{Conclusions}

We use a cosmological numerical simulation that models time-dependent
and spatially-inhomo\-geneous cosmological reionization to compute
secondary CMB temperature anisotropies in a representative CDM+$\Lambda$
cosmology.  We compare our numerical results with analytical
calculations and compliment them with the numerical simulations of
Springel et al.\ (2000) (whose mass resolution almost exactly
compliments ours, and accidentally both simulations have the same values
of cosmological parameters).  We thus are able to compute the power
spectrum of secondary anisotropies for the kinetic SZ effect and for the
patchy reionization effect over a wide range of angular scales. The
thermal SZ effect, dominated at medium $\ell$ by massive clusters of
galaxies, is more difficult to compute on all scales due to the extreme
non-linearity of the problem.

We find that the role of patchy reionization per se (i.e., evolution of
$\HII$ regions around the sources of ionization) in generating the
secondary anisotropies is subdominant. This conclusion is however only
valid when the sources of reionization are clustered on scales
comparable to the matter correlation length, in which case one can
expect to find a source (or multiple sources) in almost every dark
matter halo. If, in the opposite case, the sources of reionization are
rare (like bright quasars), the relative role of patchy reionization
will be greater, as many high density regions will remain neutral for a
long time, and may not be negligible. The latter scenario is, however,
less likely given the current observational limits on the abundances of
bright quasars (Fan et al.\ 2000b), and thus CMB observations on small
angular scales may not be as useful as hoped to study the details of
reionization.

A real window onto reionization is provided by the detailed morphology
of the fluctuations, that is, the non-Gaussianity discussed above. The
angular scale of the fluctuations traces the three-dimensional
separation scale of the ionized regions. Of course, a full
understanding of this will require not just a detection of broad-band
fluctuation power at very small angular scales, but a detailed,
high signal-to-noise map of these very small-scale fluctuations with an
angular resolution of arc seconds, which is definitely some years away.

While the details of reionization may not be easily discernible from the
broad-band power spectrum, the amplitude of the fluctuations, which is
directly related to the redshift of reionization, surely is. However, as
our comparison with Springel et al.\ (2000) shows, on medium angular
scales ($l\la10^5$) the signal is dominated by lower redshifts
($z\sim2-4$), and only on small angular scales (several arc seconds) the
amplitude of the anisotropies is related to the redshift of
reionization.  In addition, this dependence is quite weak, unless
reionization occurred close to the currently allowed observational limit
of $z_{\rm REI}\la30$. Nevertheless, observations on those angular
scales can put significant constraints on the redshift of reionization.
Such arcsecond-scale CMB observations are best accomplished with
interferometers, and indeed the recent development of compact
submillimeter-wave arrays has paved the way for such observations. For
example, the ATCA telescope finds a limit of $\langle
\ell(\ell+1)C_\ell/(2\pi)\rangle < 2\times10^{-10}$ at $\ell\sim 4000$
(Subrahmanyan et al.\ 2000), well above the values presented here. In
addition, these observations are inevitably plagued by point source
confusion, especially dangerous when the signal also has a compact
galaxy-like component.




\acknowledgements 
We would like to acknowledge helpful conversations with Marc
        Kamionkowski, Joe Silk, Andrew Liddle, Naoshi Sugiyama and
	Ue-Li Pen. AHJ
        acknowledges support from NSF KDI grant 9872979 and NASA LTSA
        grant NAG5-6552.

\placefig{\end{document}}

\clearpage

\newcounter{figurecap}
\setcounter{figurecap}{0}

\begin{center}
\bf Figure Captions
\end{center}

\refstepcounter{figurecap}
Fig.\ \thefigurecap---\label{figPB}\capPB

\refstepcounter{figurecap}
Fig.\ \thefigurecap---\label{figPS}\capPS

\refstepcounter{figurecap}
Fig.\ \thefigurecap---\label{figPC}\capPC

\refstepcounter{figurecap}
Fig.\ \thefigurecap---\label{figAC}\capAC

\refstepcounter{figurecap}
Fig.\ \thefigurecap---\label{figNC}\capNC

\refstepcounter{figurecap}
Fig.\ \thefigurecap---\label{figLT}\capLT

\refstepcounter{figurecap}
Fig.\ \thefigurecap---\label{figIM}\capIM

\refstepcounter{figurecap}
Fig.\ \thefigurecap---\label{figPD}\capPD

\refstepcounter{figurecap}
Fig.\ \thefigurecap---\label{figCC}\capCC

\refstepcounter{figurecap}
Fig.\ \thefigurecap---\label{figCO}\capCO

\refstepcounter{figurecap}
Fig.\ \thefigurecap---\label{figTT}\capTT

\refstepcounter{figurecap}
Fig.\ \thefigurecap---\label{figCV}\capCV

\refstepcounter{figurecap}
Fig.\ \thefigurecap---\label{figRV}\capRV

\refstepcounter{figurecap}
Fig.\ \thefigurecap---\label{figCL}\capCL

\refstepcounter{figurecap}
Fig.\ \thefigurecap---\label{figDZ}\capDZ


\begin{references}

\reference{Aea96}
Aghanim, N., Desert, F.\ X., Puget, J.\ L., \& Gispert, R. 1996, \aa, 311, 1

\reference{BDL00}
Baker, J.\ E., Davis, M., \& Lin, H. 2000, \apj, submitted (astro-ph/9909030)

\reference{BJ99}
Bond, J.\ R., \& Jaffe, A.\ H. 1999, Phil.\ Trans.\ Royal Soc.\ Lon.\ A, 
357, 57

\reference{BNea00}
Benson, A.\ J., Nusser, A., Sugiyama, N., Lacey, C.\ G. 2000, \mnras,
in press (astro-ph/0002457)

\reference{Bea99}
Bridle, S.\ L., et al., 1999, MNRAS, 310, 565

\reference{Bea00}
Bridle, S.\ L., Zehavi, I., Dekel. A., Lahav, O., Hobson, M.\ P.,
\& Lasenby, A.\ N. 2000, \mnras, submitted (astro-ph/0006170)

\reference{Buea00}
Bruscoli, M., Ferrara, A., Fabbri, R., \& Ciardi, B. 2000, \mnras, submitted 
(astro-ph/9911467)

\reference{dBea00}
de Bernardis, P., et al.\ 2000, \nat, 404, 955

\reference{ECF96} 
Eke, V.\ R., Cole, S., \& Frenk, C.\ S. 1996, MNRAS, 282, 263

\reference{Fea00a} 
Fan, X., et al.\ 2000a, \aj, submitted (astro-ph/0005414)

\reference{Fea00b} 
Fan, X., et al.\ 2000b, \aj, submitted (astro-ph/0008123)

\reference{FGS98}
Fardal, M.A., Giroux, M.L., \& Shull, J.M.\ 1998, \aj, 115, 2206.

\reference{G00}
Gnedin, N.\ Y. 2000, \apj, 535, 530

\reference{GBL99} 
Griffiths, L.\ M., Barbosa, D., Liddle, A.\ D. 1998, \mnras, 308, 854

\reference{GH98}
Gruzinov, A., \& Hu, W. 1998, \apj, 508, 435

\reference{HK99}
Haiman, Z., \& Knox, L. 1999, in ASP Conf.\ Ser.\ 181, Microwave Foregrounds,
ed.\ A.\ de Oliveira-Costa \& M.\ Tegmark (San Francisco: ASP), 227

\reference{Hea00}
Hanany, S., et al.\ 2000, \apjl, submitted (astro-ph/0005123)

\reference{H99}
Hu, W. 2000, \apj, 529, 12

\reference{Hea00}
Hu, W., Fukugita, M., Zaldarriaga, M., \& Tegmark, M. 2000, \apj, submitted
(astro-ph/0006436)

\reference{HW96}
Hu, W., \& White, M. 1996, \aa, 315, 33

\reference{JK98} 
Jaffe, A.\ H., \& Kamionkowski, M. 1998, Phys.\ Rev.\ D, 58, 043001

\reference{Jea00}
Jaffe, A.\ H., et al. 2000, Phys.\ Rev.\ Lett., submitted (astro-ph/0007333)

\reference{KSD98}
Knox, L., Scoccimarro, R., \& Dodelson, S. 1998, Phys.\ Rev.\ Lett., 81, 2004

\reference{M94}
Melott, A.\ L. 1994, ApJ, 426, L19

\reference{OV86} 
Ostriker, J.\ P, \& Vishniac, E. 1986, \apj, 306, 51

\reference{PJ98}
Peebles, P.\ J.\ E., \& Juszkiewicz, R. 1998, \apj, 509, 483

\reference{RKSP00}
Refregier, A., Komatsu, E., Spergel, D.\ N., \& Pen, U.-L. 2000, 
Phys.\ Rev. D, in press
(astro-ph/9912180)

\reference{SBP00}
Seljak, U., Burwell, J., \& Pen, U.-L. 2000, Phys.\ Rev. D, submitted 
(astro-ph/0001120)

\reference{dSea00}
da Silva, A.\ C., Barbosa, D., Liddle, A.\ R., \& Thomas, P.\ A. 2000, \mnras,
in press (astro-ph/9907224)

\reference{SWH00}
Springel, V., White, M., \& Hernquist, L. 2000, \apjl, submitted 
(astro-ph/0008133)

\reference{Sea00} 
Stern, D., et al.\ 2000, \apj, 533, L75

\reference{Sub00} Subrahmanyan, R., Kesteven, M. J., Ekers, R. D.,
Sinclair, M.,Silk, J., \mnras, to appear. astro-ph/0002467

\reference{TZ00} 
Tegmark, M., Zaldarriaga, M. 2000, Phys.\ Rev.\ Lett., in press 
(astro-ph/0004393)

\reference{WSP00}
White, M., Scott, D., \& Pierpaoli, E. 2000, \apj, in press (astro-ph/0004385)

\reference{Zea00} 
Zheng, W., et al.\ 2000, \aj, submitted (astro-ph/0005247)

\end{references}
\end{document}